\documentclass{svjour2} 
\smartqed
\usepackage{graphicx}                 
\usepackage{color}                    
\usepackage{hyperref}                 
\usepackage{algorithmic}

\usepackage{amsmath,amssymb}
\usepackage{a4wide}
\usepackage{bbm}
\usepackage[hang,scriptsize,nooneline]{subfigure} 
\usepackage[T1]{fontenc}
\usepackage[latin1]{inputenc}
\usepackage{floatflt}

\newtheorem{alg}{Algorithm}

\def\EE{\mathbb{E}}
\def\PP{\mathbb{P}}

\def\NN{\mathbb{N}}
\def\RR{\mathbb{R}}

\def\VV{\mathbb{V}}

\newcommand{\one}{\mathbbm 1}

\newcommand{\rd}{\mathrm{d}} 

\def\cB{\mathcal{B}}
\def\cD{\mathcal{D}}

\def\Exp{\text{\rm Exp}}
\def\U{\text{\rm Uni}}

\newcommand{\ind}{\mathbbm 1}
\title{Rare Event Simulation for T-cell activation}

\author{Florian Lipsmeier}

\begin{document}
\title{Rare event simulation for T-cell activation   
}


\author{Florian Lipsmeier        \and
        Ellen Baake
}


\institute{F.~Lipsmeier, E.~Baake \at
              Faculty of Technology, Bielefeld University, 33501 Bielefeld,  Germany\\
                                \email{\{flipsmei,ebaake\}@techfak.uni-bielefeld.de}
}

\date{Received: date / Accepted: date}

\maketitle

\begin{abstract}
The problem of \emph{statistical recognition} is considered, as it
arises in
immunobiology, namely, the discrimination of foreign antigens
against a background of the body's own molecules. The precise mechanism of 
this foreign-self-distinction, though one of the major tasks of the immune
system, continues to be a fundamental
puzzle. Recent progress has been made by van den Berg, Rand, and
Burroughs \cite{Berg:2001}, who modelled the \emph{probabilistic} nature
of the interaction between the relevant cell types, namely,
T-cells and  antigen-presenting cells (APCs). Here, the
stochasticity is due to the random sample of antigens present
on the surface of every APC, and to the random receptor type that
characterises individual T-cells. It has been shown
previously \cite{Berg:2001,Zint:2008} that this model, though highly idealised,
is capable of reproducing important aspects of the recognition phenomenon, and
of explaining them on the basis of stochastic rare events.
These results were obtained with the help of a refined large deviation
theorem and were thus asymptotic in nature. Simulations have, so far,
been restricted to the straightforward simple sampling
approach, which does not
allow for sample sizes large enough to address more detailed questions.
Building on the available large deviation results, we develop an
importance sampling technique  that allows for a
convenient exploration of the relevant tail events by means
of simulation. With its help, we investigate
the mechanism of statistical recognition in some depth. In particular,
we illustrate how a foreign antigen can stand out against the
self background if it is present in sufficiently many copies, 
although no \emph{a priori} difference between self and nonself is
built into the model.

\keywords{Immunobiology \and statistical recognition \and large deviations
\and rare event simulation}
 \PACS{87.16.af 
   \and 87.16.dr 
   \and 87.18.Tt 
   }
\subclass{92-08 
   \and   92C99 
   \and   60F10 
}
\end{abstract}


\section{Introduction}
The notion of statistical recognition between randomly encountered molecules
is central to many biological phenomena. This is particularly evident
in biological repertoires, which contain enough molecular diversity
to bind practically any randomly encountered target molecule. The receptor
repertoire of the immune system provides the best-known example
of a system displaying probability-based interactions; another one is the
olfactory receptor repertoire, which recognises multitudes of
odorants. This chance recognition is a well-established phenomenon
and has been analysed with the help of various statistical
and biophysical models; compare \cite{Lancet:1993,Rosenwald:2002}.
Here we will  tackle a model of statistical
recognition between \emph{cell surfaces} (in the sense of collections
of numerous surface molecules, rather than single ones)
of the immune system.
It describes a vital property of our immune system, which comes into
play when a virus invades the body and starts to multiply. 
Fortunately, however, sooner or later it is recognised as a foreign intruder 
by certain white blood cells, which are part of the immune system and  
start a specific immune response that finally eliminates the virus population.

\begin{figure}[ht]
\label{fig:cells}
\centering \includegraphics[width=.9\textwidth]{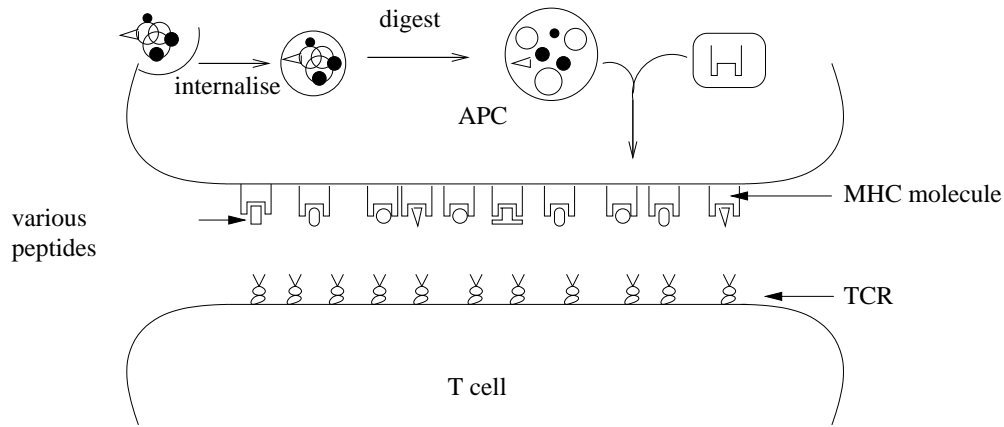}
\caption{A T-cell and an antigen-presenting cell
(based on Fig.~1 of \cite{Berg:2007}).
An {\em APC} absorbs 
molecules and particles from its vicinity 
and breaks them down. The emerging fragments, so-called 
peptides (short 
sequences of amino acids), serve as antigens.
They are bound to so-called MHC molecules (still within the cell), and
the resulting complexes, each
composed of an MHC molecule and a peptide, are 
presented on 
the surface of the cell (the MHC molecules serve as carriers or 
anchors
to the cell surface). Since most of the molecules in the vicinity of an 
APC are self molecules, every APC displays a large variety 
of different types of 
self antigens and, possibly, one (or a small number of) foreign types. The 
various 
antigen types occur in various copy numbers.
Each {\em T-cell} is characterised 
by a specific type of T-cell receptor (TCR), which is displayed in many 
{\em identical} 
copies on the surface of the particular T-cell.
When a T-cell meets an APC, the contact between them is established by a 
temporary bond between the cells,  in which the TCRs and the
MHC-peptide complexes interact with each other, which results 
in stimuli to the T-cell body.
If the added stimulation  rate 
is above a given threshold, the T-cell is
activated to reproduce, and the resulting clones of T-cells will initiate
an immune reaction against the intruder.}
\end{figure}

\begin{figure}[h]
\centering
\includegraphics[width=5.3cm,height=4.5cm]{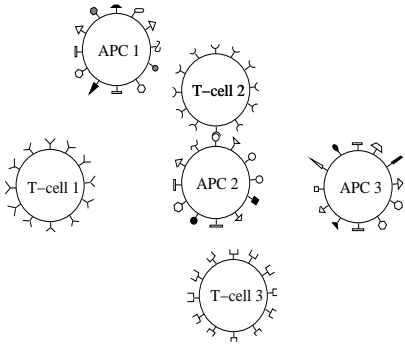}
\caption{Caricature of T-cells and APCs (from \cite{Zint:2008}). Note that 
every T-cell has many copies of one particular receptor type, but different 
T-cells have different receptor types. In contrast,  every APC carries a 
mixture of antigen types, which may appear in various copy numbers.}
\label{fig:Tcell_apc}
\end{figure}

This ability of the immune system to discriminate safely between foreign and 
self molecules is a fundamental ingredient to everyday survival of  jawed 
 vertebrates; but how this works exactly is 
still enigmatic. Indeed, the immune 
system faces an enormous 
challenge because it must recognise one (or a few) type(s) of (potentially 
dangerous) foreign molecules against an enormous variety of (harmless) self 
molecules. The particular difficulty lies in the fact that there can be 
no a priori difference between self and nonself (like some fundamental 
difference in molecular structure), since this would open up
the possibility for molecular mimicry on the part of the pathogen,
which could  quickly evolve immuno-invisibility by imitating the 
self structure.
The problem may be phrased as \emph{statistical 
recognition} of one particular foreign signal against a large, 
fluctuating self background. However, immune biology has been largely treated 
deterministically until,  recently, an explicit stochastic model was 
introduced by van den Berg, Rand and Burroughs  \cite{Berg:2001} 
(henceforth referred to as BRB) and further developed by Zint, Baake and 
den Hollander \cite{Zint:2008}. It describes (random) encounters between 
the two crucial types of white blood cells involved 
(see Figs.~\ref{fig:cells} and \ref{fig:Tcell_apc}): the 
antigen-presenting cells (APCs), which 
display a mixture of self and foreign antigens at their surface 
(a sample of the molecules around in the body), and the T-cells, which 
``scan'' the APCs by means of certain receptors and ultimately decide 
whether or not to react, i.e., to start an immune response.

To be biologically more precise, we consider the encounters 
of so-called \emph{naive T-cells}
with \emph{professional APCs} in the \emph{secondary lymphoid tissue}.
A \emph{naive T-cell} is a cell that has finished its maturation
process in the thymus and has been released into the body, where
it has not yet been exposed to antigen.
It tends to dwell in \emph{secondary lymphoid tissue} like
lymph nodes, where it comes into contact with \emph{professional APCs},
special white blood cells with so-called MHC  molecules
at their surface that serve as carriers for antigens.
Each {\em T-cell} is characterised 
by a specific type of T-cell receptor (TCR), which is displayed in many 
{\em identical} 
copies on the surface of the particular T-cell. A large number (estimated
at $10^7$ in \cite{Arstila:1999}) of different receptors, and hence
different T-cell types, are present in an individual (every type, in turn,
is present in several copies, which form a T-cell clone). However, the
number of potential antigen types is still vastly larger (roughly $10^{13}$;
see \cite{Mason:1998}). Thus, specific recognition (where one TCR
recognises exactly one antigen) is impossible; this is known as
Mason's paradox. The task is further complicated by the fact that
every APC displays on the order of thousand(s) of different self
antigen types, in various copy numbers 
\cite{Hunt:1992,Mason:1998,Stevanovic:1999}, 
together with, possibly, one (or 
a small number of) foreign types; the T-cells therefore face a literal
``needle in a haystack'' problem. 

For an encounter between a 
pair of T-cell and APC, both chosen randomly from the diverse pool of
T-cells and APCs, the probability to react must be very 
small (otherwise, immune reactions would occur permanently);
this is a central theme in the analysis. It entails that
some questions may  be answered analytically with the help 
of large deviation theory; others require simulation, but the use of this 
has been limited due to the small probabilities involved, at least with the 
straightforward simulation methods applied so far \cite{Berg:2001,Zint:2008}. 
The main purpose of this article is to devise an efficient 
importance sampling method based on large deviation theory and
tailored to the problem at hand, and to use 
this to investigate the mechanism of statistical recognition in more detail. 
The paper is organised as follows. In Sect.~2, we present the most important 
biological facts and recapitulate the model; this will be a 
self-contained, but highly simplified outline,  since the full picture is 
available elsewhere \cite{Berg:2001,Zint:2008}. In Sect.~3, we summarise 
(mainly from \cite{Dieker:2005} and \cite{Bucklew:2004}) some general theory 
that allows to design efficient methods of rare event simulation on the 
basis of a large deviation analysis, and tailor these to the problem at 
hand in Sect.~4. 
Sect.~5 presents the simulation results and analyses them both from
the computational and the biological point of view. Simulation speeds up
by a factor of nearly $1500$ relative to the straightforward approaches 
used so far. This enables us to explore regions of parameter space as yet 
inaccessible, to validate previous asymptotic results,
and to investigate the mechanism of statistical recognition 
in more depth than previously possible.

\section{The T-cell model}
\label{sec:tcell_model}
In this Section, we briefly motivate and introduce the model of T-cell recognition as first proposed by  BRB in 2001 \cite{Berg:2001} and further developed by Zint, Baake and den Hollander \cite{Zint:2008}. More precisely, we
only consider the toy version of this model, which neglects the
modification of the T-cell repertoire during maturation in the thymus.
This toy version already captures important aspects of the phenomenon
while being particularly transparent. We will come back to 
maturation (already included in \cite{Berg:2001}) in the discussion. 

When T-cells and APCs meet, the T-cell receptors bind to the various 
antigens presented by the APC \cite{Davis:2003dg}. For every single 
receptor-antigen pair, there is an association-dissociation reaction, the 
rate constants for which depend on the match of the molecular structures 
of receptor and antigen. Assuming that association is much faster than 
dissociation and that there is an abundance of receptors (so that the 
antigens are mostly in the bound state), one can describe the reaction in 
terms of the dissociation rates only.

\begin{figure}[ht]
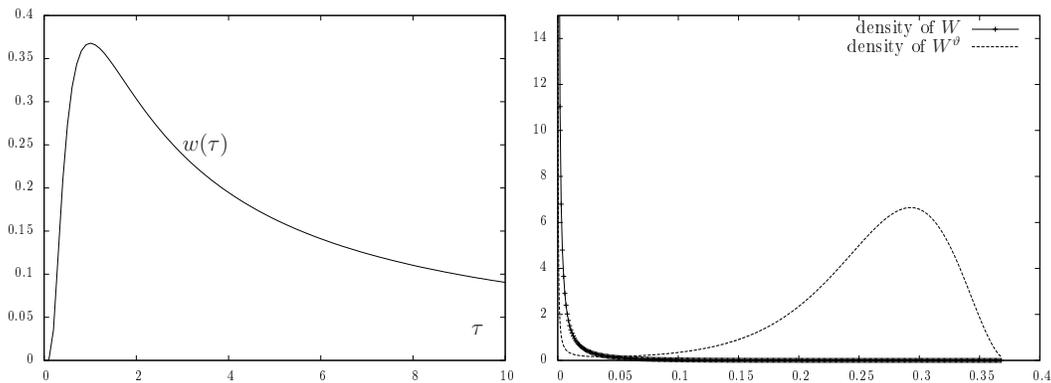

\begin{minipage}[b]{7 cm}
    \scalebox{0.6}{\input{trans_function.tex}}
  \end{minipage}
  \begin{minipage}[b]{7 cm}
    \scalebox{0.6}{\input{wdichte_tilted_tilt46.tex}}
  \end{minipage}
\caption{Left: the function $w$. Right:  the densities  of $W=w(\mathcal{T})$
and $W^{\vartheta}$ 
with tilting parameter 
$\vartheta=46$ (cf.\ Sect.~\ref{sec:res_probs}).
The densities  have poles at $w(0)=0$ and $w(1)=0.3679$ (due to the
vanishing derivative of $w$ at $\tau=0$ and $\tau=1$), but the right poles
are invisible because they support very little 
probability mass. In fact, for $\varepsilon = 0.01$, one has 
$\PP(0\leq W \leq \varepsilon) = 0.98$ and  
$\PP(w(1)-\varepsilon \leq W \leq w(1)) = 2.17 \cdot 10^{-9}$,
whereas $\PP(0\leq W^{\vartheta} \leq \varepsilon) = 0.137138$ and  
$\PP(w(1)-\varepsilon \leq W^{\vartheta} \leq w(1)) = 0.0050$.}
\label{fig:w}
\end{figure}

Every time a receptor unbinds from an antigen, it sends a signal to 
the T-cell, provided the association has lasted for at least one time unit 
(i.e., we rescale time so that the unit of time is this minimal association 
time required). The duration of a binding of a given receptor-antigen pair
follows the $\Exp(1/\tau)$ distribution, i.e.\ the exponential
distribution with mean $\tau$, where $\tau$ is the inverse dissociation rate  
of the pair in question. 
The rate of  stimuli induced  
by the interaction of our  antigen with the receptors in its 
vicinity is then given  by  
\begin{equation} 
w(\tau)= \frac{1}{\tau} \exp(-\frac{1}{\tau}),
\label{eq:transformation_function}
\end{equation} 
i.e., the dissociation rate times the probability that the 
association has lasted long enough. (If the simplifying assumption
of unlimited receptor abundance is dispensed with, 
Eq.~\eqref{eq:transformation_function} must be modified, see
\cite{Berg:2003a}.) 
As shown in Fig.~\ref{fig:w}, 
the function 
$w$ first increases and 
then decreases with $\tau$ with a maximum at $\tau=1$, which reflects the 
fact that, for $\tau<1$, the bindings tend not to last long enough, whereas 
for $\tau>1$, they tend to last so long that only few stimuli are
expected per time unit. 

The T-cell sums up the signals induced by the 
different antigens on the APC, and if the total stimulation rate
reaches a certain threshold value,
the cell initiates an immune response. This model relies on several hypotheses, 
which are known as kinetic proofreading 
\cite{McKeithan:1995cq,Rabinowitz:1996pt,Lord:1999kh,Hlavacek:2002xq}, 
serial triggering 
\cite{Valitutti:1995bc,Valitutti:1997to,Sousa:2000xu,Borovsky:2002dz,Utzny:2006hi,dushek:2008}, 
counting of stimulated TCRs \cite{Viola:1996ir,Rothenberg:1996mw},
and the optimal dwell-time hypothesis \cite{Kalergis:2001le,Gonzalez:2005bx}. 

Due to the huge amount of different receptor and antigen types, it is 
impossible (and unnecessary) to prescribe the binding durations for all 
pairs of receptor and antigen types individually. Therefore, BRB chose a 
probabilistic approach to describe the meeting of APCs and T-cells. 
A randomly chosen T-cell (that is, a randomly chosen type of receptor) 
encounters a randomly chosen APC (that is, a random mixture  of antigens). The 
mean binding time that governs the binding of this random receptor to the 
$j$th type of antigen  is taken to be a random variable denoted
by $\mathcal{T}_j$. The $\mathcal{T}_j$ are independent and identically
distributed (i.i.d.) and  
are assumed to follow the ${\rm Exp}(1/\bar{\tau})$ distribution,
i.e., the exponential distribution with mean $\bar{\tau}$,
where $\bar{\tau}$ is a free parameter. 
Note that there are two exponential distributions (and two levels
of averaging) involved here. First, the duration of an \emph{individual
binding} between a type-$j$ antigen
and a random receptor is ${\rm Exp}(1/\mathcal{T}_j)$ distributed
(see the discussion of Eq.~\eqref{eq:transformation_function}). Second,
$\mathcal{T}_j$, the \emph{mean duration} of  such a binding
(where the receptor is chosen  once and
the times are averaged over repeated bindings with a $j$ antigen) 
is itself an exponential random variable,
with realisation $\tau_j$. Finally, its mean,
$\EE(\mathcal{T}_j)=\bar \tau$, is the mean binding time of a $j$-antigen
(and, due to the i.i.d.\ assumption, of any antigen)
when averaged over all encounters with
the various receptor types. The exponential distribution of the
individual binding time is an immediate consequence of the
(first-order) unbinding kinetics. In contrast, the corresponding
assumption for the $\mathcal{T}_j$ is made for simplicity; 
the approach is compatible
with various other distributions as well,
see \cite{Berg:2001} and \cite{Zint:2008}. The i.i.d.\ assumption, however,
is crucial, since it implies, in particular, that
there is no difference between self and foreign antigens here; i.e., no a 
priori distinction is built into the model. 

The total stimulation a T-cell 
receives is the sum over all stimulus rates $W_j=w(\mathcal{T}_j)$ that emerge
from antigens of the $j$'th type. It is further assumed that there is at most 
one type of foreign antigen in $z^{(f)}$ copies on an APC, whose signal must be 
discriminated against the signals of a huge amount of self antigens.
(There could, in principle, be multiple foreign peptide types,
but there are good reasons to assume that there are mechanisms
to ensure that a given T-cell sees at
most one foreign peptide type, see \cite{Berg:2003a}). The self 
antigens are here divided into two distinct classes, $c$ and $v$, 
that are present in 
different copy numbers $z^{(c)}$ and $z^{(v)}$. An APC displays  
$m^{(c)}$ and $m^{(v)}$ 
different types of class $c$ and $v$. The indices $c$ and $v$ stand for 
constitutive and for variable, respectively;
 but for the purpose of this article,
only the abundancies are relevant, in particular, $z^{(c)} > z^{(v)}$
and $m^{(c)} <m^{(v)}$.
Over the whole APC the total number of antigens 
is then $m^{(c)} z^{(c)} + m^{(v)}  z^{(v)} =: M$ if no foreign antigen
is present. If $z^{(f)}$ foreign molecules are also present, the self
molecules are assumed to be proportionally displaced (via the factor
$q := (M - z^{(f)})/M$), so that the total number of antigens remains
unchanged at
\begin{equation}
z^{(f)}+m^{(c)} q z^{(c)} + m^{(v)} q z^{(v)} = M.
\label{eq:total_n}
\end{equation}

The total stimulation rate in  a random encounter of T-cell and APC
can then be described as a function of  $z^{(f)}$:
\begin{equation}
G(z^{(f)}) := \left ( \sum_{j=1}^{m^{(c)}} q z^{(c)} W_j \right )
+ \left ( \sum_{j=m^{(c)}+1}^{m^{(c)}+m^{(v)}} q z^{(v)} W_j \right ) + z^{(f)} W_{m^{(c)}+m^{(v)}+1},
\label{eq:the_model}
\end{equation}
i.e., a weighted sum of i.i.d.\ random variables.
Alternatively, we consider the extension of the model proposed by 
Zint et al. \cite{Zint:2008}, which, instead of the deterministic copy 
numbers $z^{(c)},z^{(v)}$, uses random variables $Z_j^{(c)},Z_j^{(v)}$ 
distributed according to binomial distributions with 
$\mathbb{E}(Z_j^{(c)})=z^{(c)} \, , \, \mathbb{E}(Z_j^{(v)})=z^{(v)}$,
where $\mathbb{E}$ denotes expectation
(so the expected number of antigens per APC is still $M$). The model
then reads
\begin{equation}
G(z^{(f)}): = 
\left ( \sum_{j=1}^{m^{(c)}} q Z_j^{(c)} W_j \right )
+ \left ( \sum_{j=m^{(c)}+1}^{m^{(c)}+m^{(v)}} q Z_j^{(v)} W_j \right )
+ z^{(f)} W_{m^{(c)}+m^{(v)}+1}.
\label{eq:alt_model}
\end{equation}

In line with \cite{Berg:2001,Zint:2008}, we numerically specify
the model parameters as follows:  $\bar \tau=0.04$; $m^{(c)}=50$, $m^{(v)}=1500$,
$z^{(c)} = 500$, $z^{(v)}=50$ (and hence $M=10^5$). The  distributions in the 
extended model are the binomials
${\rm Bin}(\zeta^{(c)},p)$ and ${\rm Bin}(\zeta^{(v)},p)$ for $Z_j^{(c)}$ and $Z_j^{(v)}$ 
respectively, where  $\zeta^{(c)}=1000,\, \zeta^{(v)}=100,\, {\rm and}\, p=0.5$.

The relevant quantity for us is now the probability 
\begin{equation}
\mathbb{P}\big ( G(z^{(f)})\geq g_{\rm act} \big )
\end{equation}
that the stimulation rate reaches or surpasses
a threshold $g_{\rm act}$. To achieve a good 
foreign-self discrimination, there must be a large difference in probability 
between the stimulation rate in the case with self antigens only ($z^{(f)}=0$), 
and the stimulation rate with the foreign antigen present, i.e.,
\begin{equation}
1 \gg \mathbb{P}\big (G(z^{(f)}  )\geq g_{\rm act} \big )
\gg \mathbb{P} \big (G(0)\geq g_{\rm act} \big ) \geq 0 
\label{eq:bed}
\end{equation}
for realistic values of $z^{(f)}$.
Note that both events must be rare events -- otherwise, the immune
system would ``fire'' all the time. Thus $g_{\rm act}$ must be much larger than 
$\mathbb{E}(G(z^{(f)}))$ (which, due to  \eqref{eq:total_n} and the identical 
distribution of the $W_j$, is independent of $z^{(f)}$). 
Evaluating these small probabilities is a challenge.
So far, two routes have been used: analytic (asymptotic) 
theory based on large deviations (LD) and straightforward simulation 
(so-called simple sampling). Both have their shortcomings: the LD approach 
is only 
exact in the limit of infinitely many antigen types (and the available
error estimates are usually too crude to be useful);
the simulation strategy, on the other hand, 
is so time-consuming that it becomes simply impossible to obtain sample 
sizes large enough for
a detailed analysis, in particular for
large values of $g_{\rm act}$. Therefore, an importance sampling approach 
is required. 
Let us now recapitulate some underlying theory.

\section{Rare event simulation: general theory}
\label{sec:resim}
The general problem we now consider is to estimate the probability 
$P(A)$ of a (rare) event $A$ under a probability measure $P$. The 
straightforward approach, known as simple sampling, uses the estimate 
\begin{equation}\label{eq:ss}
(\widehat{P(A)})_N := \frac{1}{N}\sum_{i=1}^N \ind\{S^{(i)} \in A\}
= \frac{1}{N} {\rm card} \{1 \leq i \leq N \mid S^{(i)} \in A \},
\end{equation}
where the $\{ S^{(i)} \}_{1\leq i \leq N}$ are independent and identically
distributed (i.i.d.) random variables with distribution $P$, $\one\{.\}$ 
denotes the indicator function, and $N$ is the
sample size;
we will throughout use $\widehat v$ for an estimate of a quantity $v$.
$(\widehat{P(A)})_N$
is obviously an unbiased and consistent estimate, but, for small $P(A)$, the convergence  to $P(A)$ is slow, and large samples are required to get reliable estimates.

Various simulation methods are available that deal with this problem 
and yield  a  better rate of convergence 
(see the monograph by Bucklew \cite{Bucklew:2004} for an overview). 
Most of them achieve this improvement by reducing the variance of the 
estimator. We will concentrate here on the most wide-spread class of
methods, namely importance sampling. As is well known, one introduces 
a new sampling distribution $Q$ here under which $A$ is more likely to happen, produces samples from this distribution and returns  to the original distribution by  reweighting.
In general, finding a good importance sampling distribution that
reduces the variance as much as possible is an art, and much of
the literature revolves around this. Some
general purpose and many ad hoc strategies exist, but usually, 
importance sampling
distributions are best tailored by exploiting the structure of the
specific problem at hand. However, if the problem can be embedded
into a sequence of problems for which a so-called large deviation principle
is valid, a unified theory is available
that identifies the most efficient simulation distribution.
This technique of ``large deviation simulation'' was introduced by Sadowski
and Bucklew \cite{Sadowsky:1990}, laid down in the
monograph by Bucklew \cite{Bucklew:2004}, and further developed by
Dieker and Mandjes \cite{Dieker:2005}. It rests on the well-established
theory of large deviations, as summarised, for example, in the books
by Dembo and Zeitouni \cite{Dembo:1998} or den Hollander \cite{Hollander:2000}.
Let us recapitulate the basic
background.

\subsection{Large deviation probabilities}
\label{sec:large_dev_probs}
Consider a sequence $\{S_n\}$ of random variables on the probability
space $(\RR^d, \cB, \PP)$, where $\cB$ is the Borel $\sigma$-algebra
of $\RR^d$. Let $\{P_n\}$ be the family of probability measures
induced by $\{S_n\}$, i.e., $P_n(B) = \PP(S_n \in B)$ for
$B \in \cB$. We assume
throughout that $\{S_n\}$ satisfies a large deviation principle (LDP)
according to the following definition \cite{Dembo:1998,Dieker:2005}:

\begin{definition}[Large deviation principle]\label{def:LDP}
A family of probability measures $\{P_n\}$ on $(\RR^d, \cB)$ 
satisfies the large deviation principle (LDP) with rate function 
$I$ if $I: \RR^d \to [0, \infty]$ is lower semicontinuous 
and, for all $B \in \cB$,
\begin{equation}
  - \inf_{x \in B^{\circ}} I(x) 
  \leq \liminf_{n \to \infty} \frac{1}{n} \log P_n(B)
  \leq \limsup_{n \to \infty} \frac{1}{n} \log P_n(B)
  \leq - \inf_{x \in \overline B} I(x),
\end{equation}
where $B^{\circ}:= {\rm int} (B)$ and $\overline B:= {\rm clos}(B)$ denote 
the interior and the closure of 
$B$, respectively. $I$ is said to be a good rate function
if it has compact level sets in that 
$I^{-1}([0,c]) = \{x \in \RR^d: I(x) \leq c\}$ is compact for all
$c \in \RR^d$. \qed
\label{def:ldp}
\end{definition}

A set $B$ is called an $I$-\emph{continuity set} if
\begin{equation}\label{eq:contset}
\inf_{x \in B^{\circ}} I(x) = \inf_{x \in B} I(x) = \inf_{x \in \overline B} I(x).
\end{equation}
If $B$ is such a set, the LDP means that $P_n(B)$ decays 
exponentially for large $n$,
with decay coefficient $\inf_{x \in B} I(x)$.
A point $b$  is called a \emph{minimum rate point} of $B$
if $\inf_{x \in B} I(x)=I(b)$.


Large deviation principles are well known for many families of
random variables, like empirical means of i.i.d.\ random variables
or empirical measures of Markov chains. For the application we have
in mind, which involves sums of independent, but not identically distributed
random variables, we need the fairly general setting of the
G\"artner-Ellis theorem, which we recapitulate here  
(cf.\ \cite[Thm.~2.3.6]{Dembo:1998} and \cite[Ch.~V]{Hollander:2000}). 
Let $\varphi_n(\vartheta) := \EE_{P_n}(e^{\langle \vartheta, S_n \rangle})$,
$\vartheta \in \RR^d$, be the moment-generating function of $S_n$, where
$\langle .,.\rangle$ denotes the scalar product and
$\EE_{\mu} (.)$ denotes the expectation of a random variable
with respect to the probability measure $\mu$.

\begin{theorem}[G\"artner-Ellis]
\label{gaertner_ellis}
Assume that
\begin{enumerate}
\item[\rm(G1)] 
$  \lim_{n \to \infty} \frac{1}{n} \log \varphi_n(n \vartheta) 
  =: \Lambda(\vartheta) \in [-\infty, \infty]  \, \text{exists},
$
\item[\rm(G2)]
$
  0 \in {\rm int}(\cD_{\Lambda}), \quad \text{where} \;
  \cD_{\Lambda} := \{ \vartheta \in \RR^d: \Lambda(\vartheta) < \infty \}$
is the \emph{effective domain} of  $\Lambda$,
  \item[\rm(G3)] $\Lambda$ is lower semi-continuous on $\RR^d$,
   \item[\rm(G4)] $\Lambda$ is differentiable on ${\rm int} (\cD_{\Lambda})$,
   \item[\rm(G5)] Either $\cD_{\Lambda} = \RR^d$ or $\Lambda$ is
                  steep at its boundary $\partial \cD_{\Lambda}$,
                  i.e., 
                  $\lim_{{\rm int}(\cD_{\Lambda}) \ni \vartheta 
                  \to \partial \cD_{\Lambda}} | \nabla \Lambda(\vartheta) | = \infty$.
\end{enumerate}
Then, $\{P_n\}$ satisfies the LDP on $\RR^d$ with good
rate function $I$,
where $I$ is the Legendre transform of $\Lambda$, i.e.,
\begin{equation}\label{eq:I}
  I(x) = 
  \sup_{\vartheta \in \RR^d} [ \langle x, \vartheta \rangle - \Lambda(\vartheta)],
  \quad x \in \RR^d.
\end{equation}
\qed
\end{theorem}

The function $\Lambda$ in (G1) is convex. If there is a solution $\vartheta^*$
of 
\begin{equation}\label{eq:theta_a}
     \nabla \Lambda(\vartheta) = x,
\end{equation}
one has
\begin{equation}\label{eq:Ix}
I(x) = \langle \vartheta^*,x \rangle - \Lambda(\vartheta^*).
\end{equation}
If $\Lambda$ is strictly convex in all directions, $\vartheta^*$ is
unique. See Fig.~\ref{fig:ratefunc} for a one-dimensional example
(the T-cell application, in fact).

\subsection{Simulating rare event probabilities}
\label{sec:res_probs}
Let now $A \in \cB$ be a \emph{rare event}  
in the sense that $0<\inf_{x \in A} I(x) <\infty$. 
Here, the
first inequality  implies that
$A$ becomes exponentially unlikely as $n \to \infty$,
whereas the second inequality serves to exclude nongeneric cases
(in particular cases where the event is  impossible).
An important notion for the rare event simulation of $P_n(A)$
is that of a \emph{dominating point} \cite[p.~83]{Bucklew:2004}:
A point 
  $a$ is a \emph{dominating point} of the set $A$ if it is the unique point 
such that

  {\rm a)} $a \in \partial A$,

  {\rm b)} $\exists$ a unique solution $\vartheta^*$ of 
  $\nabla \Lambda(\vartheta) = a$, and 
  
  {\rm c)} $A \subset \{x \in \RR^d:
  \langle \vartheta^*, x-a \rangle \geq 0\}$.

A dominating point, if it exists, is always a unique minimum rate
point (see \cite[p.~83]{Bucklew:2004}). 
Convexity of $A$  implies existence of a dominating point
(cf.~\cite{Dieker:2005}).

Following \cite{Dieker:2005} we now turn to the
problem of simulating $P_n(A)=  \EE_{P_n}(\one\{S_n \in A\})$.
The naive simple-sampling
estimate obtained from $N$ i.i.d.\ copies $S_n^{(i)}$
($1 \leq i \leq N$), drawn from
$P_n$, is, as in \eqref{eq:ss}, given by
\begin{equation}
   \big (\widehat{P_n(A)} \big )_N := \frac{1}{N} \sum_{i=1}^N \one\{S_n^{(i)} \in A\}.
\label{eq:simple_sampling}
\end{equation}
It is unbiased and converges (almost surely)
to $P_n(A)$ in the limit $N \to \infty$,
but it is inefficient since it requires that $N$
increase exponentially with $n$ to yield a meaningful estimate.
Instead of $\{S_n\}$, one therefore considers an alternative family
of random variables, $\{T_n\}$  
with  distribution family $\{Q_n\}$, again  on $(\RR^d, \cB)$, 
under which $A$ occurs more frequently. Assuming that 
$P_n$ and $Q_n$ are absolutely continuous with respect to each
other, one can use the
identity
\begin{equation}
  P_n(A) =  \EE_{P_n}(\one\{S_n \in A\}) = \EE_{Q_n} \Big (\one\{T_n \in A\}
   \frac{\rd P_n}{\rd Q_n} (T_n) \Big ),
\end{equation}
where ${\rm d}P_n/{\rm d}Q_n$ is the Radon-Nikodym derivative of $P_n$
with respect to $Q_n$. The resulting 
importance sampling estimate then relies on i.i.d.\ samples
$T_n^{(i)}$ from $\{Q_n\}$ and reads 
\begin{equation}
   \big (\widehat{P_{Q_n}(A)} \big )_N := 
   \frac{1}{N} \sum_{i=1}^N \one\{T_n^{(i)} \in A\} 
   \frac{\rd P_n}{\rd Q_n} (T_n^{(i)}) , 
\label{eq:rewe}
\end{equation}
where $({\rm d}P_n/{\rm d}Q_n)(.)$ acts as a reweighting factor from the sampling 
distribution to the original one. It is reasonable to assume that
$({\rm d}P_n/{\rm d}Q_n)$ is continuous  to avoid the usual problems
with $L^1$-functions; this is no restriction for our intended
application.

An adequate optimality concept in this context is that of
\emph{asymptotic efficiency}. According to \cite{Dieker:2005},
it is based on the \emph{relative error} $\eta^{}_N(Q_n,A)$
defined via its square
\begin{equation}\label{eq:eta}
  \eta_N^2(Q_n,A) := \frac{\VV_{Q_n} \big (\widehat{P_{Q_n}(A)} \big )_N}
  {\big (P_n(A) \big )^2}
\end{equation}
(where $\VV_{\mu}(.)$ denotes the variance of a random variable
with respect to the probability measure $\mu$).
The relative error is proportional to the width of the confidence
interval relative to the (expected) estimate itself. Asymptotic efficiency
is then defined as follows.
\begin{definition}[Asymptotic efficiency]
  An importance sampling family $\{Q_n\}$ is called
  \emph{asymptotically efficient} for the rare event $A$ if
\begin{equation}
\label{eq:asym_eff}
   \lim_{n \to \infty} \frac{1}{n} \log N_{Q_n}^* = 0,
\end{equation}
  where $N_{Q_n}^* := 
  \inf\{N \in \NN: \eta^{}_N(Q_n,A) \leq \eta^{}_{\rm max} \}$
  for some given maximal relative error $\eta^{}_{\rm max}$,
  $0 < \eta^{}_{\rm max} < \infty$.
\end{definition}
In words, asymptotic efficiency means that the number of samples required to keep the relative error below a prescribed bound $\eta^{}_{\rm max}$ increases only subexponentially (rather than exponentially as with simple sampling).
The concrete choice of $\eta^{}_{\rm max }$ is actually irrelevant, see
Lemma 1 in \cite{Dieker:2005}.

An obvious idea from large deviation theory would be to use, as
sampling distributions, the family of measures $\{P_n^{\vartheta}\}$
that are exponentially tilted with parameter $\vartheta$, that is,
\begin{equation}
  \frac{\rd P_n^{\vartheta}}{\rd P_n} (x) = 
  \frac{e^{n \langle \vartheta, x \rangle}}{\varphi_n(n\vartheta)}, 
\quad x \in \RR^d;
\end{equation}
$P_n^{\vartheta}$ then takes the role of $Q_n$.
The task remains to find a suitable $\vartheta$, i.e., a 
tilting parameter that makes $\{P_n^{\vartheta}\}$ asymptotically efficient.
Necessary and sufficient conditions for this are given
in \cite[Assumption 1 and Corollary 1]{Dieker:2005} 
and are summarised below,
in a form adapted to the present context.

\begin{theorem}[Dieker-Mandjes 2005]
\label{thm:varadhan}
Assume that, for some given $\vartheta^*$,
\begin{enumerate}
 \item[{\rm (V1)}] $\{P_n\}$ satisfies an LDP with good rate function $I$,
 \item[{\rm (V2)}] $\limsup_{n \to \infty} \frac{1}{n} 
             \log \varphi_n(\gamma n \vartheta^*) < \infty$ for some $\gamma > 1$,
             and, likewise, with $\vartheta^*$ replaced by $- \vartheta^*$,
 \item[{\rm (V3)}] The rare event $A$ is both an $I$-continuity set and an 
            $(I+\langle \vartheta^*, . \rangle)$-continuity set. 
\end{enumerate}
Then, the tilted measure $\{P_n^{\vartheta^*}\}$ 
is asymptotically efficient for simulating $A$ if and only if
\begin{equation}\label{eq:vara}
\inf_{x \in \RR^d} [I(x) - \langle \vartheta^*,x \rangle ]
             + \inf_{x \in \overline A} [I(x) + \langle \vartheta^*,x \rangle ]
             = 2 \inf_{x \in A^{\circ}} I(x).
    \end{equation}
\end{theorem}

We use assumption (V2) here  to replace the weaker but
less easy to verify condition (2) in Assumption~1 of \cite{Dieker:2005},
in line with the paragraph below (2) in \cite{Dieker:2005}, or
\cite[Thm.~4.3.1]{Dembo:1998}.
Note also that (V2) holds automatically if $\varphi_n(n \vartheta)$
exists for all $\vartheta$ -- but this is not mandatory here, since
only a given $\vartheta^*$ is considered. 

The proof of Theorem~\ref{thm:varadhan}  is given in \cite{Dieker:2005}
and need not be recapitulated here; but we would like to
comment briefly  on what happens in the central
condition \eqref{eq:vara}. Replacing $Q_n$ by $P_n^{\vartheta^*}$
in \eqref{eq:eta} and \eqref{eq:rewe}, we can rewrite
$\eta_N^2$ as
\begin{equation}\label{eq:explain_vara}
\begin{split}
  \eta_N^2(P_n^{\vartheta^*},A) 
& = \frac{\VV_{P_n^{\vartheta^*}}(\widehat{P_{P_n^{\vartheta^*}}(A)})_N}
{\big ( P_n(A)\big)^2} 
 = \frac{1}{N} \frac{\VV_{P_n^{\vartheta^*}}(\widehat{P_{P_n^{\vartheta^*}}(A)})_1}{\big ( P_n(A)\big)^2} \\
& = \frac{1}{N} \frac{1}{\big ( P_n(A)\big)^2}
  \Big [\int_A \Big 
     (\frac{\rd P_n}{\rd P_n^{\vartheta^*}} \Big )^2 \rd P_n^{\vartheta^*} - \big (P_n(A) \big )^2 \Big ].
\end{split}
\end{equation}
Obviously (by (V1) and (V3)), $ 2 \inf_{x \in A^{\circ}} I(x)$ (i.e., the right-hand side of
\eqref{eq:vara}) is the exponential decay rate of $(P_n(A))^2$.
Inspection of the proof of Theorem~\ref{thm:varadhan} reveals that
the left-hand side of \eqref{eq:vara} is the exponential decay rate of 
$\int_A \Big (\frac{\rd P_n}{\rd P_n^{\vartheta^*}} \Big )^2 \rd 
P_n^{\vartheta^*}$.
It is clear from \eqref{eq:explain_vara} that, for asymptotic efficiency
to hold, $\int_A \Big (\frac{\rd P_n}{\rd P_n^{\vartheta^*}} \Big )^2 
\rd P_n^{\vartheta^*}$ must
tend to zero at least as fast as  $(P_n(A))^2$. But it cannot
decrease faster,  since 
$\VV_{P_n^{\vartheta^*}}(\widehat{P_{P_n^{\vartheta^*}}(A)})_1$ 
is nonnegative, so that 
$\int_A \Big (\frac{\rd P_n}{\rd Q_n} \Big )^2 \rd Q_n \geq (P_n(A))^2$
for arbitrary $Q_n$. Hence, the exponential decay rates must be exactly
equal, as stated by \eqref{eq:vara}. (A closely related argument is given
in \cite[Ch.~5.2]{Bucklew:2004}.)

Theorem \ref{thm:varadhan} is widely applicable. It holds in many
standard situations, in particular in many of those that arise
in applications.

\begin{proposition}\label{prop:GEV3}
  Let $\{P_n\}$ be a family of probability measures that satisfy
  the conditions of the G\"artner-Ellis theorem, with (good) rate 
function $I$. Let  $A$ be a rare event  with dominating point $a$,
let  $\vartheta^*$ be the unique solution
of $\nabla \Lambda(\vartheta)=a$, and assume {\rm (V2)} and {\rm (V3)}.
Then $\{P_n^{\vartheta^*}\}$ is the unique
  tilted family that is asymptotically efficient for 
  simulating $P_n(A)$.
\end{proposition}

\begin{proof}
  The proof is a simple application of Thm.~\ref{thm:varadhan}.
  (V1) follows from the G\"artner-Ellis theorem;
  we only need to verify condition \eqref{eq:vara}.
    For the first infimum in \eqref{eq:vara}, one obtains
   \begin{equation}\label{eq:first_inf}
    \inf_{x \in \RR^d} [ I(x) - \langle \vartheta^*, x \rangle ]
    = - \Lambda(\vartheta^*) = I(a) - \langle \vartheta^*, a \rangle.
  \end{equation}
  Here, the first step follows from the convex duality lemma 
  (compare \cite[Lemma 4.5.8]{Dembo:1998}), 
  which is applicable since $\Lambda$ is lower semicontinuous by
  (G3), and convex and $> - \infty$ everywhere (this follows from (G1) and (G2)
  by \cite[Lemma V.4]{Hollander:2000}). The second step 
  is due to part b) of the dominating point property of $a$,
   together with Eq.~(\ref{eq:Ix}).
 
  As to the second infimum in \eqref{eq:vara},
  $a$ minimises both $I$ and $\langle \vartheta^*,.\rangle$ on $A$
(by the dominating point property).  
Together with (V3), this gives
  \begin{equation}\label{eq:second_inf}
    \inf_{x \in \overline A} [ I(x) + \langle \vartheta^*, x \rangle ]
    = I(a) + \langle \vartheta^*, a \rangle.
  \end{equation}
Eqs.~\eqref{eq:first_inf} and \eqref{eq:second_inf} together give
\eqref{eq:vara} because $\inf_{x \in A^{\circ}} 
I(x) = \inf_{x \in  \partial A} I(x)
=I(a)$.  
\qed
\end{proof}

\begin{remark}
Note that an efficiency result closely related to Proposition \ref{prop:GEV3}
has previously been given by
Bucklew \cite[Thm.~5.2.1]{Bucklew:2004}, but this is based on the
variance rather than the relative error; and it is only a sufficient
condition.

Note also that our assumption of a dominating point greatly simplifies
the situation. Theorem~2 also allows to
cope with situations without a dominating point -- but this is not needed
below.
\end{remark}

Let us now apply this theory to  the T-cell model.

\section{Rare event simulation: the T-cell model}

Recall that simulating the T-cell model means sampling the random variables $G(z^{(f)})$ of \eqref{eq:the_model} and estimating the corresponding tail probabilities $\mathbb{P}(G(z^{(f)})\geq g_{\rm act})$. Inspection of Eq.~\eqref{eq:the_model} reveals two difficulties:
\begin{enumerate}
\item $G(z^{(f)})$ is a weighted sum of i.i.d.\ random variables, to which 
the standard results for sums of i.i.d.\ random variables (in particular, 
Cram\'{e}r's theorem) are not applicable. We therefore need an extension 
to weighted sums -- or, better, to general sums of independent, but not 
identically distributed random variables, which include weighted sums as 
a simple special case. This 
is straightforward and will be the subject of Sect.~\ref{sec:ld_simulation}. 
In particular, it will be seen that, like in the 
i.i.d.\ case, every term in the sum must be tilted with the same parameter, 
but now this global tilting factor is a function of all the individual 
distributions involved.

\item Simulating the random variables $W_j = w(\mathcal{T}_j)$ is 
straightforward via simple sampling: draw ${\rm Exp}(1/\bar{\tau})$ distributed 
random numbers $\tau_j$ (as realisations of $\mathcal{T}_j$) and apply the 
transformation \eqref{eq:transformation_function}. However, simulating the 
corresponding tilted variables is a difficult task, for two reasons. First 
of all, there is no indication of how to sample from the tilted 
distribution via 
transformation of one  of the elementary distributions (like $\U_{[0,1]}$
(the uniform distribution on the unit interval), 
or ${\rm Exp}(\lambda ))$ for which efficient random number generation is 
possible. Although such a transformation might exist in principle, there 
is no systematic way of finding it. One
reason for this is that tilting acts at the level of the densities, 
but even the original (untilted) density of $W=w(\mathcal{T})$ is not 
available explicitly. (With $W$ and $\mathcal{T}$ (without indices) we
mean any representative of the family.)
This is because its calculation requires the 
inverse functions and derivatives of the two branches 
(increasing and decreasing) of the function $w$,
but these are unavailable analytically.

In the absence  of a transformation method, one might consider to determine the 
tilted density numerically, integrate it (again numerically) and discretise 
and tabulate the resulting distribution function. However, this is, again, 
forbidding for our particular function $w$: due to the vanishing 
derivatives at $\mathcal{T}=0$ and $\mathcal{T}=1$, the transformation 
formula for densities
yields singularities in the density of $W$  at these values, with a
sizeable fraction of the probability mass concentrated very close to $0$ (see
Fig.~\ref{fig:w}). This renders numerical calculations unreliable.
To circumvent these problems, we will, in Sect.~\ref{sec:new_tilting}, present a sampling method for the tilted random variable $W^{\vartheta}$ that is based on tilting $\mathcal{T}$ rather than $W$ itself.

\end{enumerate}

\subsection{Large deviations for independent but not identically distributed
random variables}
\label{sec:ld_simulation}

We consider $K$ independent families of i.i.d.\  $\RR^d$-valued 
random variables, $\{Y_{\ell}^{(1)}\},$  $\ldots,\{Y_{\ell}^{(K)}\}$
(i.e., the distribution within any given family $\{Y_{\ell}^{(k)}\}$,
$1 \leq k \leq K$, is fixed,
but the distributions may vary across families). 
Assume that  
$\Lambda^{(k)}(\vartheta) := 
\log \EE(e^{\langle \vartheta, Y_1^{(k)} \rangle })$, 
the log moment-generating function of $Y_1^{(k)}$,
is finite for all $\vartheta \in \RR^d$ and $1 \leq k \leq K$
(here,  $\EE(.)$  refers
to the probability measure induced by the random variable involved).
Let $n^{(1)}, \ldots, n^{(K)}$ be positive integers,  $n:= \sum_{k=1}^K n^{(k)}$,
\begin{equation}
  V_n := \sum_{\ell=1}^{n^{(1)}} Y_{\ell}^{(1)} + \ldots +  
  \sum_{\ell=1}^{n^{(K)}} Y_{\ell}^{(K)},
\label{eq:sum}
\end{equation}
and $P_n$ be the probability measure induced by $S_n=V_n/n$.
In the limit $n \to \infty$, subject to $n^{(k)}/n \to \gamma^{(k)}$ for all
$1 \leq k \leq K$, the limiting log-moment generating function
of $\{S_n\}$  becomes
\begin{equation}\label{eq:weighted_Lambda}
  \Lambda(\vartheta) = 
  \lim_{n \to \infty} \frac{1}{n} \log \EE(e^{\langle \vartheta, V_n \rangle})
  = \lim_{n \to \infty} \sum_{k=1}^K  \frac{n^{(k)}}{n} \Lambda^{(k)}(\vartheta) 
  = \sum_{k=1}^K  \gamma^{(k)} \Lambda^{(k)}(\vartheta),
\end{equation}
where the second step is due to independence.
Since, by assumption, $\Lambda^{(k)}(\vartheta)<\infty$ for all 
$\vartheta \in \RR^d$ and $1 \leq k \leq K$,
the $\Lambda^{(k)}$ are differentiable on all of $\RR^d$
(see \cite[Lemma 2.2.31]{Dembo:1998}); in fact, they are
even $C^{\infty}(\RR^d)$ \cite[Ex.ercise 2.2.24]{Dembo:1998}.  
Thus, $\Lambda$ is $C^{\infty}(\RR^d)$ as well.

By \eqref{eq:weighted_Lambda}, we have (G1). 
Again due to $\Lambda^{(k)}(\vartheta)<\infty$,
(G2) and (G5) are automatically satisfied. Furthermore,
the differentiability of $\Lambda$ entails (G3) and (G4). 
We have therefore shown

\begin{lemma}\label{lemma:i.ni.d.}
Under the  assumptions of this paragraph, $\{P_n\}$ 
satisfies the G\"artner-Ellis theorem,
with rate function $I$ given by Eq.~\eqref{eq:I}. \qed
\end{lemma}
Such $\{P_n\}$ are therefore candidates for efficient simulation
according to Prop.~\ref{prop:GEV3}.
The tilting factor $\vartheta^*$ may not be accessible analytically, but
can be evaluated numerically from \eqref{eq:theta_a}. Due to independence, tilting of $S_n$ with
$n\vartheta^*$ (that is, tilting of $V_n$ with $\vartheta^*$)
is equivalent to tilting each $Y_{\ell}^{(k)}$ with $\vartheta^*$.

\subsection{Tilting of transformed random variables}
\label{sec:new_tilting}
Unlike the $W_j$, the 
${\rm Exp}(1/\bar{\tau})$-distributed random variables $\mathcal{T}_j$ are 
tilted easily (tilting with  $\vartheta$ 
simply gives ${\rm Exp}(-\vartheta + 1/\bar{\tau})$). One is therefore tempted 
to tilt the $\mathcal{T}_j$ rather than the $W_j$, or, in other words, to interchange
the order of tilting and transformation. The following Theorem states
the key idea.

\begin{theorem}
\label{fact:new_tilting}
Let $X$ be an $\mathbb{R}^d$-valued random variable with 
probability measure $\mu$, and let $Y:= h \circ X$ (or $Y=h(X)$ by
slight abuse of notation), where
$h: \RR^d \rightarrow \RR^d$ is $\mu$-measurable.
Then $Y$ has probability measure $\nu=\mu \circ h^{-1}$,
where $h^{-1}(y)$ denotes the preimage of $y$. Assume now that
$\mathbb{E}_{\mu}(e^{ \langle\vartheta ,h(X)\rangle})$ exists,
let  $\tilde{X}^{\vartheta}$ be an $\mathbb{R}^d$-valued random variable
with probability measure $\tilde \mu^{\vartheta}$ related to $\mu$
via 
\begin{equation}
\frac{{\rm d} \tilde \mu^{\vartheta}}{{\rm d} \mu} (x) 
= \frac{e^{\langle \vartheta, h(x) \rangle}}{\mathbb{E}_{\mu}(e^{\langle \vartheta, h(X) \rangle})}
\label{eq:tilt_density}
\end{equation}
(so that $\tilde \mu^{\vartheta} \ll \mu$),
and let $\tilde Y^{\vartheta}=h(\tilde X^{\vartheta})$.  
Then, the measures $\tilde \nu^{\vartheta}$ (of $\tilde Y^{\vartheta}$)
and $\nu^{\vartheta}$ (for the tilted version of $\nu$,  belonging
to $Y^{\vartheta}$) are equal,
where $\nu^{\vartheta} \ll \nu$ with Radon-Nikodym density
\begin{equation}
\frac{\rd  \nu^{\vartheta}}{\rd \nu}(y) = 
\frac{e^{\langle\vartheta ,y \rangle }}{\mathbb{E}_{\nu}(e^{\langle \vartheta ,Y \rangle }) }.
\label{eq:tilting_frac}
\end{equation}
\end{theorem}
\begin{proof}
Note first that $e^{\langle \vartheta,y \rangle}$ is clearly $\mu$-measurable,
and
\begin{equation}
\mathbb{E}_{\nu}(e^{\langle \vartheta ,Y \rangle }) =
\int_{\RR^d} e^{\langle \vartheta,y \rangle} \rd \nu(y) 
=  \int_{\RR^d} e^{\langle \vartheta,h(x) \rangle} \rd \mu(x)
= \mathbb{E}_{\mu}(e^{\langle \vartheta ,h(X) \rangle }), 
\end{equation}
which exists by assumption, so $\nu^{\vartheta}$ is well-defined.
We now have to show that $\tilde \nu^{\vartheta}(B) = \nu^{\vartheta}(B)$
for arbitrary Borel sets $B$. Observing that 
$\tilde \nu^{\vartheta} = \tilde \mu^{\vartheta} \circ h^{-1}$ and
employing the formulas for transformation of measures \cite[(13.7)]{Bill:1995}
and change of variable \cite[Thm.~16.13]{Bill:1995},
together with \eqref{eq:tilt_density}, one indeed
obtains
\begin{equation}
\begin{split}
 \tilde \nu^{\vartheta}(B) &= \tilde \mu^{\vartheta}\big (h^{-1}(B) \big ) 
 = \int_{h^{-1}(B)} \frac{\rd \tilde \mu^{\vartheta}}{\rd \mu} (x) \rd \mu (x) 
 = \frac{1}{\mathbb{E}_{\mu}(e^{\langle \vartheta ,h(X) \rangle })}
 \int_{h^{-1}(B)} e^{\langle \vartheta,h(x) \rangle} \rd \mu(x) \\
& = \frac{1}{\mathbb{E}_{\nu}(e^{\langle \vartheta ,Y \rangle })}
\int_{B} e^{\langle \vartheta,y \rangle} \rd \nu(y) 
= \int_{B} \frac{\rd  \nu^{\vartheta}}{\rd \nu} (y) \rd \nu (y) 
 = \nu^{\vartheta}(B),
\end{split}
\end{equation}
which proves the claim. \qed
\end{proof}
In words, Theorem~\ref{fact:new_tilting}  is nothing but 
the simple observation that, to obtain the tilted version of $Y=h(X)$,
one can reweight the measure $\mu$ of $X$ with the 
factors $e^{\langle \vartheta, h(x)\rangle}$, rather than reweighting 
the measure $\nu$ of $Y$ with $e^{\langle \vartheta, y\rangle}$.
It should be clear, however, that the  measure
$\tilde{\mu}^{\vartheta}$ differs from the usual tilted version of 
$\mu$, which  would involve  tilting factors 
$e^{\langle \vartheta, x\rangle}$ rather than 
$e^{\langle \vartheta, h(x)\rangle }$; for this reason, we use the 
notation $\tilde{\mu}^{\vartheta}$ rather than $\mu^{\vartheta}$.
Such kind of tilting is  common
in large deviation theory (see, e.g., \cite[Chap.~2.1.2]{Dembo:1998}).
Nevertheless, the 
simple observation above is the key to simulation 
if $\mu$ (and $\tilde{\mu}^{\vartheta}$) 
are readily 
accessible at least numerically, but $\nu$ (and $\nu^{\vartheta}$) are not.

This is precisely our situation, with 
$\tilde{\mathcal{T}}^{\vartheta}$, $\alpha W^{\vartheta}$ and $\alpha w$
($\alpha \in \{q z^{(c)},q z^{(v)},z^{(f)}\}$), respectively, taking
the roles of 
$\tilde{X}^{\vartheta}$, $Y^{\vartheta}$ and $h$ (we will  use 
$f$, $\tilde{f}^{\vartheta}$, $g$ and $g^{\vartheta}$ for the corresponding 
densities of $\mathcal{T}$, $\tilde{\mathcal{T}}^{\vartheta}$, $\alpha W$, and
$(\alpha W)^{\vartheta}$). Still, reweighting of the exponential density of 
$\mathcal{T}$ with 
$e^{\vartheta \alpha  w(\tau)}$
does not yield an explicit closed-form density, 
and no direct simulation method is available for the corresponding
random variables. However, the reweighted densities are easily accessible 
numerically, in contrast to those of $W$ and its tilted variant, 
$W^{\vartheta}$. 
The problem may thus be solved by  calculating and integrating 
$\tilde{f}^{\vartheta}$ numerically and discretising and tabulating the
resulting distribution function $\tilde{F}^{\vartheta}$. Samples of 
$\tilde{\mathcal{T}}^{\vartheta}$ may then be drawn according to this table 
(i.e., by formally looking up the solution of 
$\tilde{F}(\tilde{\mathcal{T}}^{\vartheta})=U$ 
for $ U\sim {\rm Uni}_{[0,1]}$), and 
$\alpha W^{\vartheta}= \alpha w(\tilde{\mathcal{T}}^{\vartheta})$ is then readily evaluated. 
The only difficulty left is the time required for searching the table. 
But this is a practical matter and will be dealt with in the next 
paragraph.

\subsection{The algorithm}\label{subsec:alg}
Taking together our theoretical results, we can now detail the specific 
importance sampling algorithm for the simulation of the T-cell model of 
Sect. \ref{sec:tcell_model}. If not stated otherwise, we will refer
to the basic model \eqref{eq:the_model}. Recall that it 
describes the stimulation rate $G(z^{(f)})$
and we wish to evaluate the probability 
$\mathbb{P}(G(z^{(f)})\geq g_{\rm act})$.

To apply LD sampling, let us embed the model into a sequence of models 
with increasing total number $n=n^{(c)}+n^{(v)}+n^{(f)}$ of antigen types, 
where $n^{(c)}$, $n^{(v)}$, 
and $n^{(f)}$ are the numbers of constitutive, variable and foreign
antigen types. (This is an aritificial sequence of models required
to formulate the limiting process involved in the theory; in contrast
to the original model, there can now be multiple foreign antigen
types.)
Let
\begin{equation}
G_n(z^{(f)}) = \left( \sum_{j=1}^{n^{(c)}} q_n z^{(c)} W_j \right)+ \left( \sum_{j=n^{(c)}+1}^{n^{(c)}+n^{(v)}} q_n z^{(v)} W_j \right) +  \left ( \sum_{j=n^{(c)}+n^{(v)}+1}^{n^{(c)}+n^{(v)}+n^{(f)}}z^{(f)} W_{n^{(c)}+n^{(v)}+j} \right ),
\end{equation}
where
\begin{equation}
q_n = \frac{n^{(c)} z^{(c)} + n^{(v)} z^{(v)} - n^{(f)} z^{(f)}}{n^{(c)} z^{(c)} + n^{(v)} z^{(v)}}
\end{equation}
(where $z^{(c)}$, $z^{(v)}$, and $z^{(f)}$ are independent of $n$).
Clearly, $G_n(z^{(f)})$ coincides with $G(z^{(f)})$ of \eqref{eq:the_model}
if $n^{(c)}=m^{(c)}$, $n^{(v)}=m^{(v)}$, and $n^{(f)}=m^{(f)}$, where 
$m^{(f)}=0$ or $m^{(f)}=1$ 
depending on whether $z^{(f)}=0$ or $z^{(f)}>0$; then, 
$n=m=m^{(c)}+m^{(v)}+m^{(f)}$.
We have to consider
$\PP \big ( G_n(z^{(f)})/n > g_{\rm act} /m \big )$
(this reflects the fact that
$g_{\rm act}$ must scale with system size).
The sequences $\{G_n(z^{(f)})\}$ and  $\{G_n(z^{(f)})\}/n $ take
the roles of $\{V_n\}$ and  $\{S_n\}$, respectively, in Secs. 
\ref{sec:large_dev_probs} and \ref{sec:ld_simulation},
with $P_n$ the law of $G_n(z^{(f)})/n$; and we consider 
$A=[g_{\rm act}/m,\infty)$ with 
$\EE( G_m(z^{(f)})/m) < g_{\rm act}/m < Mw(1)/m$ (the latter is
the maximum value of $ G_m(z^{(f)})/m$ since $w(\tau)$ has its
maximum at $\tau=1$).
The limit $n \rightarrow \infty$ is then taken so that 
$\lim_{n \rightarrow \infty} n^{(c)}/n = m^{(c)}/m$,
$\lim_{n \rightarrow \infty} n^{(v)}/n = m^{(v)}/m$, as well as
$\lim_{n \rightarrow \infty} n^{(f)}/n = m^{(f)}/m$,
that is, the relative amounts of constitutive, variable, and foreign
antigens  approach those fixed in the original model,
\eqref{eq:the_model}. (Note that, in \cite{Zint:2008}, a different
limit was employed, namely,  $n \rightarrow \infty$ with
$\lim_{n \to \infty} n^{(c)}/n^{(v)} = C_1 \in (0,\infty)$ and 
$\lim_{n \to \infty} n^{(f)}/n = 0$; this is appropriate for exact
asymptotics, but not for simulation, because the asymptotic tilting
factor  to be used in the latter then does not 
feel the foreign antigens.)

\begin{figure}[ht]
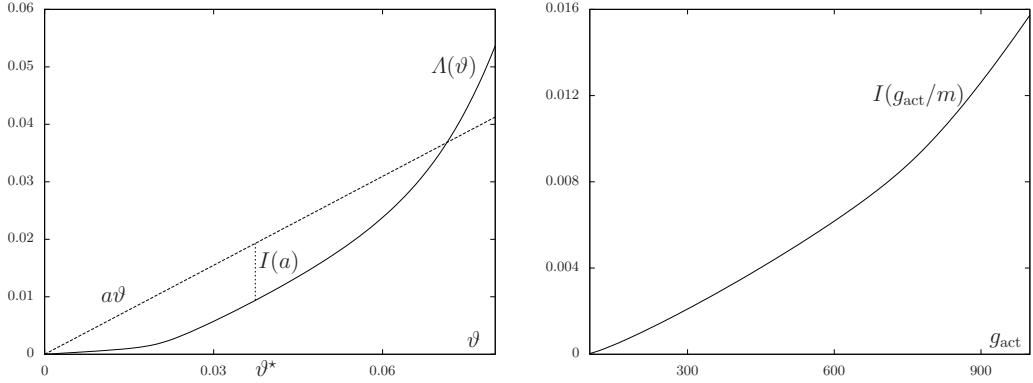

\begin{minipage}[b]{7 cm}
    \scalebox{0.6}{\input{cumulant_function.tex}}
  \end{minipage}
  \begin{minipage}[b]{7 cm}
    \scalebox{0.6}{\input{ratefunction.tex}}
  \end{minipage}
\caption{The cumulant-generating function $\Lambda$  (left) 
and the
rate function $I$ (right) for the T-cell model \eqref{eq:the_model}.
The slope of the straight line in the left panel is $a=g_{\rm act}/m$,
where $g_{\rm act}=800$ and $m=1551$. At $\vartheta^*$, 
$a \vartheta- \Lambda(\vartheta)$ assumes its maximum, $I(a)$
(cf.\ \eqref{eq:I} --\eqref{eq:Ix}).}
\label{fig:ratefunc}
\end{figure}

\begin{lemma}
Let $f$ be the density of $\Exp(1/\overline \tau)$ (i.e.,
$f(\tau)= e^{-\tau/\overline \tau}/\overline \tau$), and
\begin{equation}
\psi(t): = \EE(e^{tW}) = \int_{0}^{\infty} \exp \big (t w(\tau) \big ) f(\tau) d\tau
=\frac{1}{\bar{\tau} } \int_{0}^{\infty} \exp \left(t \frac{\exp(-1/\tau )}{\tau}-\frac{\tau}{\bar{\tau}}\right) d\tau
\end{equation}
be the moment-generating function of $W_1$. 
Under the assumptions of Sect.~{\rm \ref{subsec:alg}}, 
the unique solution $\vartheta^*$ of
\begin{equation}
\begin{split}
\frac{g_{\rm act}}{m} & =  \frac{m^{(c)}}{m} q z^{(c)} \left. 
\left[ \frac{\rd}{\rd t} 
\log \psi(t)\right]\right|_{t=q z^{(c)} \vartheta} 
+\frac{m^{(v)}}{m} q z^{(v)} \left. \left[ \frac{\rd}{\rd t} \log \psi(t)\right]\right|_{t=q z^{(v)} \vartheta}\\
&+ \frac{1}{m} z^{(f)} \left. \left[ \frac{\rd}{\rd t} \log \psi(t)\right]\right|_{t=z^{(f)} \vartheta}
\end{split}
\label{eq:solve_tilting}
\end{equation}
is the unique asymptotically efficient tilting parameter for LD
simulation of $P_n(A)$.
\end{lemma}
\begin{proof}
Clearly, $P_n$ satisfies the assumptions of Sect.~\ref{sec:ld_simulation}.
Note, in particular, that $\psi(t) < \infty$ 
for all
$t \in \RR$ since $W$ is bounded above and below, and so  
\begin{equation}
\Lambda(\vartheta) = \lim_{n \to \infty} \log \EE(e^{\vartheta  G_n(z^{(f)})/n})
=  \frac{m^{(c)}}{m} \log \psi(q z^{(c)} \vartheta) + \frac{m^{(v)}}{m} \log  \psi(q z^{(v)} \vartheta)
   +  \frac{1}{m} \log \psi(z^{(f)} \vartheta) < \infty
\label{eq:Lambda_T}
\end{equation}
for all $\vartheta$; hence, the G\"artner-Ellis theorem holds by
Lemma~\ref{lemma:i.ni.d.}.
To verify the remaining assumptions of Prop.~\ref{prop:GEV3},
recall from Sec.~\ref{sec:ld_simulation} that 
$\Lambda(\vartheta)$ is differentiable (with continuous derivative)
on all of $\RR$. 
The bounds on $g_{\rm act}/m$ lead to
\begin{equation}\label{eq:derivs}
 \Lambda'(0) = \frac{\EE \big ( G(z^{(f)} \big )}{m} <
 \frac{g_{\rm act}}{m} <  \frac{Mw(1)}{m} = \lim_{\vartheta \to \infty} \Lambda'(\vartheta). 
\end{equation}
$\Lambda$ is strictly convex (since $(\rd^2/\rd t^2) \log \psi(t)$
is the variance of $W^t$, the tilted version of $W$
(cf.\ \cite[Prop.~XII.1.1]{Asmussen:2003}),
which is positive since $W$ and hence $W^t$ is nondegenerate).
Eq.~\eqref{eq:derivs} thus entails that  $\Lambda'(\vartheta)= g_{\rm act}/m$
has a unique solution $\vartheta^*$, which is positive (and clearly
satisfies (V2)). As a consequence, $g_{\rm act}/m$ is a dominating point
of $A$, which is a rare event since $0 < I(g_{\rm act}/m)< \infty$
(by  $\Lambda(0)=0$ together
with \eqref{eq:derivs} and \eqref{eq:Ix}; cf.\ Fig.~\ref{fig:ratefunc}, left).
Finally, $A$ is a continuity set of both $I$ and $I + \langle \vartheta^*,.\rangle$
simply because $I$ and $\langle \vartheta^*,.\rangle$ are
continuous  at $g_{\rm act}/m$, and $A=\overline{A^{\circ}}$.
Realising that the right-hand side of \eqref{eq:solve_tilting}
equals $\Lambda'(\vartheta)$ (see also Eq.~(20) in \cite{Zint:2008}), 
one obtains
the claim  from Prop.~\ref{prop:GEV3}. \qed
\end{proof}

The  solution of \eqref{eq:solve_tilting} is readily 
calculated numerically. The function $\Lambda$, and
the resulting rate function $I$, are shown in Fig.~\ref{fig:ratefunc}.

As described in Sect.~\ref{sec:new_tilting}, we now tilt the density $f$ 
of the $\mathcal{T}_j$ with $\vartheta^*$
according to Eq.~\eqref{eq:tilt_density}. 
This yields three different densities $\tilde f^{\vartheta^*}_{\alpha}$,
depending on the weighting  factors $\alpha \in \{q z^{(c)}\, , \, q z^{(v)} \, , \, z^{(f)}\}$, namely
\begin{equation}
\tilde{f}_{\alpha}^{\vartheta^*}(\tau)=
\frac{\exp(\alpha \vartheta^* w(\tau)) f(\tau)}{\psi(\alpha \vartheta^*)} 
= \frac{\frac{1}{\bar{\tau} }  \exp \left( \alpha \vartheta^* \frac{\exp(-1/\tau )}{\tau}-\frac{\tau}{\bar{\tau}}\right)}{\psi(\alpha \vartheta^*) }.
\label{eq:tilted_densities}
\end{equation}
As discussed in Sect.~\ref{sec:new_tilting}, this is not the density of any known standard distribution (let alone an exponential one), and simulating from it requires numerical integration (which is well-behaved since 
the $\tilde{f}^{\vartheta^*}_{\alpha}$ are numerically well-behaved), and discretisation and
tabulation of the resulting distribution functions 
$\tilde{F}^{\vartheta^*}_{\alpha}$, followed by looking up the solution 
$\tilde \tau^{\vartheta^*}$ of 
$\tilde{F}^{\vartheta^*}_{\alpha}(\tilde{\mathcal{T}}^{\vartheta^*})=U$ for $U \sim {\rm Uni}_{[0,1]}$, to finally yield $\alpha W^{\vartheta^*}$ via 
$\alpha W^{\vartheta^*}=\alpha w(\tilde{\mathcal{T}}^{\vartheta^*})$. 

Searching the table would be the speed- (or precision-) limiting step, 
requiring $\mathcal{O}(\log D)$ operations if $D$ is the number of 
discretisation steps. This can be remedied by applying the so-called 
\emph{alias method} to quickly generate random variables according to the 
discretised probability distribution. For a description of the method, we refer the reader to \cite[pp.~25--27]{Madras:2002},
\cite{Kronmal:1979}, or 
\cite[p.~248]{Ross:2002}. Let us just summarise here that, after a 
preprocessing step, which is  done once for a given distribution, the
method only requires one $\U_{[0,1]}$ random variable together
with one multiplication, one cutoff and one subtraction 
(or two $\U_{[0,1]}$ 
random variables together with one multiplication, one cutoff and
one comparison, depending
on the implementation) to generate one 
realisation of $\tilde{\mathcal{T}}^{\vartheta^*}$, regardless of $D$ 
(in particular, it does without
searching altogether).

We now have everything at hand to formulate the  algorithm
to simulate (realisations of) $G(z^{(f)})$ of \eqref{eq:the_model}.
(For notational convenience, we will not distinguish between
random variables and their realisations here).
\begin{alg}
\begin{algorithmic}
\STATE 

\STATE compute $\vartheta^*$ by solving Eq. \eqref{eq:solve_tilting}
numerically
\STATE calculate the tilted densities  $\tilde{f}_{\alpha}^{\vartheta^*}$,
$\alpha \in \{q z^{(c)},q z^{(v)},z^{(f)}\}$, via
\eqref{eq:tilted_densities}
\FOR{i=1 till sample size N} 
\STATE for every summand $j$ of \eqref{eq:the_model} generate a sample 
$(\tilde{\mathcal{T}}_j^{\vartheta^*})^{(i)}$ according to its density 
$\tilde{f}^{\vartheta^*}_{\alpha(j)}$ with the help of the alias method 
(here, the upper index $(i)$ is added to reflect sample $i$, and
$\alpha(j)$ is the weighting factor of the sum to which   $j$ belongs)
\STATE calculate  
\[
  \big (G(z^{(f)})\big )^{(i)} = \Bigg ( \sum_{j=1}^{m^{(c)}} q z^{(c)} w\big ((\tilde{\mathcal{T}}^{\vartheta^*}_j)^{(i)}\big ) \Bigg ) + 
\Bigg ( \sum_{j=m^{(c)}+1}^{m^{(c)}+m^{(v)}} q z^{(v)} w \big ((\tilde{\mathcal{T}}_j^{\vartheta^*})^{(i)} \big ) \Bigg ) + z^{(f)} w \big ((\tilde{\mathcal{T}}_{m^{(c)}+m^{(v)}+1}^{\vartheta^*})^{(i)} \big )
\]
\STATE calculate the indicator function times the reweighting factor (i.e.,
the $i$-th summand in Eq.~\eqref{eq:rewe})
\IF{$(G(z^{(f)}))^{(i)}\geq g_{\rm act}$}
\STATE 
$\displaystyle{
R^{(i)} = \prod_{j=1}^{m}  \frac{f_{\alpha(j)}((\tilde{\mathcal{T}}_j^{\vartheta^*})^{(i)})}{\tilde f_{\alpha(j)}^{\vartheta^*}((\tilde{\mathcal{T}}_j^{\vartheta^*})^{(i)})}
}$
\ELSE
\STATE $R^{(i)}=0$ 
\ENDIF
\ENDFOR
\STATE calculate 
$\displaystyle{
\big( \widehat{P_{P_m}^{\vartheta^*} (A)}\big)_N 
= \frac{\sum_{i=1}^{N} R^{(i)}}{N},
}$
as estimate of 
$\mathbb{P}(G(z^{(f)})>g_{\rm act})$.
\end{algorithmic}
\label{ref:alg}
\end{alg}

\subsection{Extension to variable copy numbers}
\label{subsec:ext_copy}
Let us now consider the extended model \eqref{eq:alt_model}, in which
the copy numbers are themselves random variables. This is also
covered by the large deviation theory presented above; in particular,
Lemma \ref{lemma:i.ni.d.} again applies if
the $Y^{(k)}_{\ell}$ in \eqref{eq:sum} are identified with $Z^{(c)}_j W_j$ or 
$Z^{(v)}_j W_j$, respectively. The global tilting factor
$\vartheta^*$ is, in the usual way, calculated as the solution
of $\Lambda'(\vartheta)=g_{\rm act}/m$, where $\Lambda(\vartheta)$
is as in \eqref{eq:Lambda_T} with $\psi(q z^{(k)}\vartheta)
=\EE(e^{q z^{(k)}\vartheta W})$ replaced by $\EE(\psi(q Z^{(k)}\vartheta))
=\EE(e^{q Z^{(k)}\vartheta W})$, $k \in \{c,v\}$;
see Eq.~(20) in \cite{Zint:2008}. 

However, the object of tilting now is the \emph{joint distribution}
of $W$ and $Z^{(c)}$ (or $Z^{(v)}$, respectively), that is,
${\rm d}F(\tau) {\rm d}H^{(k)}(z)$ receives the reweighting factor
$\exp(q \vartheta z w(\tau))$, where $F$ and $H^{(k)}$ denote
the measures of $\mathcal{T}$ and $Z^{(k)}$, $k \in \{c,v\}$,
respectively. This introduces dependencies between
copy numbers and stimulation rates. The resulting \emph{bivariate}
simulation task is costly and may offset some of the efficiency
gain obtained by tilting. 

If, however, the $Z^{(k)}$ are closely peaked around their means
(as is the case for our choice of parameters), the following hybrid
procedure turns out to be both practical and fast: Draw
the $Z^{(k)}$ from their original (untilted, binomial) distributions;
and simulate a tilted version of $qW$, denoted by 
$(\overline{q W})^{\vartheta^*}$, by reweighting the original
density of $qW$ with $\exp(q \vartheta^* \EE(Z^{(k)}) W)$,
irrespective of the actual value of $Z$. Clearly, this method
is not asymptotically efficient, but it is a valid importance
sampling method that turns out to compare well with
the ideal procedure used for the fixed copy numbers (see 
Sec.~\ref{subsec:error}).

\section{Results}

Let us now present the results of our simulations in two steps.
We  first investigate the performance of the  method, and then 
use it to gain more insight into
the underlying phenomenon of statistical recognition.

\subsection{Performance of the simulation method}
We will examine the performance of the importance-sampling method in three
 respects: we will compare it to simple sampling (the previously-used 
simulation method) and to the results of exact asymptotics (the 
previously-used analytic method); finally, we will quantify the efficiency 
in terms of the relative error (and thus return to the theory of 
Sect.~\ref{sec:res_probs}). In any case, we will consider 
$\mathbb{P}(G(z^{(f)})\geq g_{\rm act})$  as a function of $g_{\rm act}$ 
(and for various values of the parameter $z^{(f)}$). Of course,
this probability is just one minus
the distribution function of $G(z^{(f)})$; in immunobiology, the corresponding 
graph is known as the activation curve. 

Evaluating this graph by LD simulation 
requires, for each value of $g_{\rm act}$ to be considered, a fresh sample, 
simulated with its individual tilting factor $\vartheta^*$
(recall that this depends on $g_{\rm act}$ via \eqref{eq:solve_tilting}).
At first sight, this looks like an enormous disadvantage relative to 
simple sampling, where no threshold needs to be specified in advance; 
rather, the outcomes of the simulation  directly yield
an estimate over the entire range of the
activation curve. However, it will turn out that this
disadvantage is offset many times by the specific efficiency of hitting the rare events in LD sampling. 
(There is room for improvement: the samples that do
not hit a given rare event could be used to improve  the estimates of the more
likely events.)

\subsubsection{Comparison with simple sampling}
Clearly, both the simple-sampling and the importance-sampling estimates are unbiased and converge to the true values as $N \rightarrow \infty$. It is therefore no surprise that they yield practically identical results wherever they can be compared -- and this yields a first quick consistency check for our method. 
\begin{figure}[h]
\input{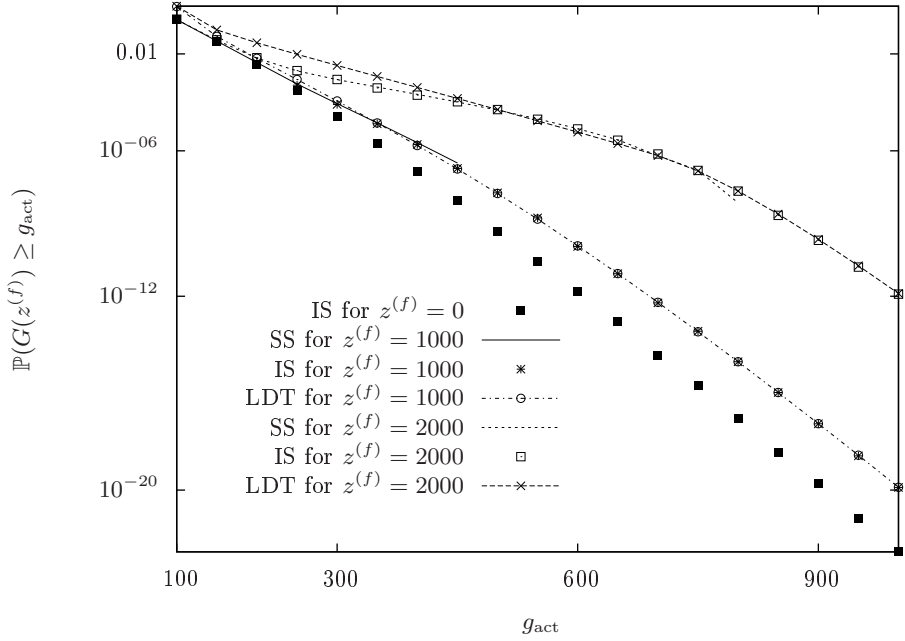}

\caption{Estimates of  the activation curve, 
$\mathbb{P}(G(z^{(f)})\geq g_{\rm act})$, in the basic
model \eqref{eq:the_model} for $z^{(f)}=1000$ and $z^{(f)}=2000$, as 
well as for the self background ($z^{(f)}=0$), on logarithmic scale. 
The probabilities were estimated independently
 with simple sampling (SS), importance sampling (IS),
and exact asymptotics based on large deviation theory (LDT) as 
used  in \cite{Zint:2008}.   For  IS, 
$19$  values of $g_{\rm act}$ were considered
(from $100$ to $1000$ in steps of $50$), and 
$N=10000$ samples were generated for each value (i.e., $1.9*10^5$ 
samples altogether),
whereas for the SS simulation,
$N=1.3*10^8$ samples were used over the entire range. The SS curves end at
$g_{\rm act}=400$ and $g_{\rm act}=800$, respectively,
because larger values were not hit in the given sample.
The IS and SS graphs agree perfectly 
until the SS simulation lacks precision. For larger threshold values,
we see a perfect agreement of the IS and LDT graphs. Note the general 
feature that, for threshold values that are not too small, the activation 
probability in the presence of foreign antigens is several orders of 
magnitude larger than the self background, i.e. Eq.~\eqref{eq:bed} 
is satisfied.}
\label{fig:vergleich}
\end{figure}

This is demonstrated in Fig.~\ref{fig:vergleich}, which shows simple 
sampling (SS) and importance sampling (IS) activation curves, each for 
$z^{(f)}=1000$ and $z^{(f)}=2000$. For SS, $ N=1.3*10^8$ samples, 
$(G(z^{(f)}))^{(i)}, 1 \leq i \leq N$,
were generated altogether for every graph, whereas for IS, 
$N=10000$ samples were generated for every threshold value considered 
(from $g_{\rm act}=100$ to $g_{\rm act}=1000$ in steps of $50$), i.e. 
$1.9*10^5$ samples altogether. 
Beyond $g_{\rm act}=450$ and $g_{\rm act}=800$ 
(for $z^{(f)}=1000$ and $z^{(f)}=2000$, respectively), no  estimates 
could be obtained via SS due to the low probabilities involved, whereas with 
IS, it is easy to get beyond $g_{\rm act}=900$ in either case, although the 
probabilities can get down to $10^{-20}$ (note,
however, that this far end of the distribution is no longer biologically
relevant). In terms of 
runtime, determining an activation curve (over its entire range) by
SS took 48 hours of CPU time (Intel Pentium M 1.4 GHz 512MB 
RAM), whereas IS required only about 2 minutes (in the threshold regime 
where the methods are comparable), that is, a speedup by a factor of
nearly 1500 is achieved. 

\begin{figure}[h]
\input{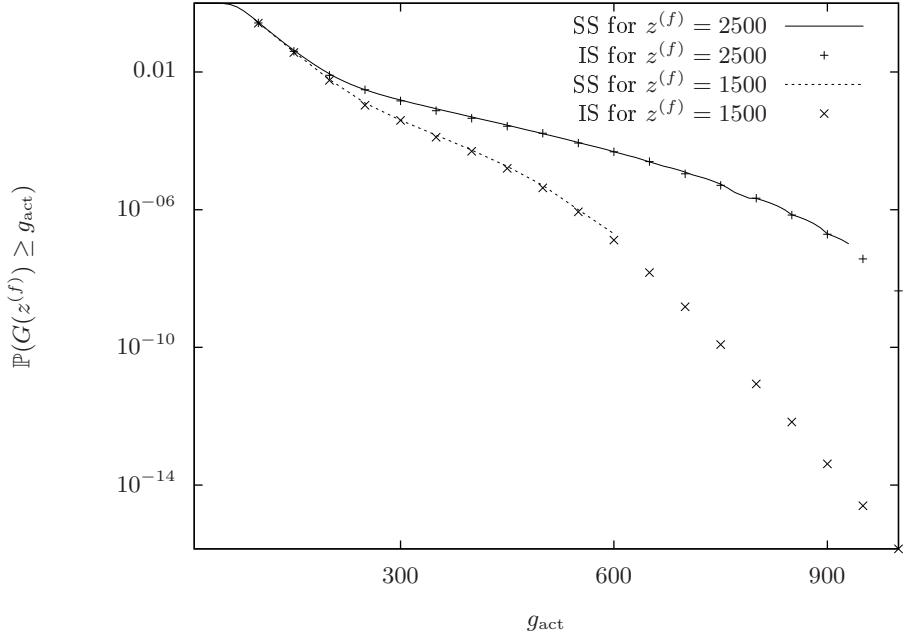}
\caption{Simulation  of  $\mathbb{P}(G(z^{(f)})\geq g_{\rm act})$ 
in the extended model \eqref{eq:alt_model},
for $z^{(f)}=1500$ and $z^{(f)}=2500$. The probabilities were estimated 
independently with simple sampling, and with importance sampling at $19$ 
different threshold values (from $100$ to $1000$ in steps of $50$).
For IS,  $9.5* 10^4$ samples were generated ($5000$ per threshold); for 
SS, $2*10^7$ samples were used.
No  estimates are obtained
with SS  at thresholds beyond 600 or 920, respectively, in analogy with
the situation in Fig.~\ref{fig:vergleich}.}
\label{fig:extended_model}
\end{figure}

We also applied our method to the extended model \eqref{eq:alt_model} with
binomially distributed copy numbers. 
Figure \ref{fig:extended_model} shows the simulation results for two
values of $z^{(f)}$, each for SS and IS. Again, the curves agree, as they must. As to runtime, it took about $130$ hours to generate the $2*10^7$ samples for SS, 
whereas for IS it took 10 min.\  to generate the $9.5* 10^4$ samples.

\subsubsection{Comparison with exact asymptotics}
A pillar of the previous analysis of Zint et al. \cite{Zint:2008} (and its 
precursor BRB \cite{Berg:2001}) has been so-called exact asymptotics. This 
is a refinement of large deviation theory which yields estimates for the 
probabilities $P_n(A)$ themselves, rather than just their exponential decay 
rates obtained via the LDP in Def.~\ref{def:ldp}. With standard large 
deviation theory (and our simulation method), it shares the tilting parameter 
which is calculated according to Eq. \eqref{eq:solve_tilting}; for more 
details, we refer to \cite{Zint:2008}. A comparison of IS 
simulation with exact asymptotics  is also included in 
Fig.~\ref{fig:vergleich}. For small values of $g_{\rm act}$, exact asymptotics 
is slightly imprecise. This is due to the asymptotic nature 
($n\rightarrow \infty$) of the method, which yields more precise results 
in the very tail of the distribution, where the deviations are
truly large. Note that, although our tilting 
factors agree with those in exact asymptotics, rare event simulation 
does not suffer from this accuracy problem since, due to the reweighting, 
it is always a valid importance sampling scheme that yields unbiased estimates 
for every finite $n$; the finite-size effects will only manifest themselves 
as a certain loss of efficiency, as will be seen below.

\subsubsection{Asymptotic efficiency and relative error}
\label{subsec:error}

In order to investigate the relative error of 
$(\widehat{P_{P_n^{\vartheta^*}}(A)})_N$, we first note that
the variance of the estimator is given by
\begin{equation}
\mathbb{V}\Big ( \big (\widehat{P_{P_n^{\vartheta^*}} (A)} \big )_N \Big )
= \frac{1}{N} \mathbb{V}\Big ( \big (\widehat{P_{P_n^{\vartheta^*}}(A)} \big)_1 
\Big )
= \frac{1}{N} \mathbb{E} \Big [ \Big ( \ind\{(T_n^{\vartheta^*})^{(1)} 
\in A\} \frac{\rd P}{\rd P_n^{\vartheta^*}} \big ( (T_n^{\vartheta^*})^{(1)}
\big )  -  P_n(A) \Big )^2 \Big ],
\label{eq:var}
\end{equation}
where we have used \eqref{eq:rewe} for $N=1$. 
$\mathbb{V}\big ( (\widehat{P_{P_n^{\vartheta^*}}(A))}_1 \big )$
can be estimated via the given number $N$ of samples in a single simulation
run, i.e., as the sample variance
\begin{equation}
\widehat{\mathbb{V}}
\Big ( \big (\widehat{P_{P_n^{\vartheta^*}}(A)} \big )_1 \Big )
= \frac{1}{N-1}  \sum_{i=1}^N \Big ( \ind\{(t_n^{\vartheta^*})^{(i)} 
\in A\} \frac{\rd P}{\rd P_n^{\vartheta^*}} 
\big ( (t_n^{\vartheta^*})^{(i)} \big )
-  \big (\widehat{P_{P_n^{\vartheta^*}}(A)} \big )_N \Big )^2,
\label{eq:emp_var}
\end{equation}
where the $(t_n^{\vartheta^*})^{(i)}$ are now considered as realisations of  
$(T_n^{\vartheta^*})^{(1)}$. We can thus estimate the squared relative error 
as
\begin{equation}
  \widehat{\eta_N^2}(P_n^{\vartheta^*},A) = \frac{1}{N} 
  \frac{\widehat{\mathbb{V}}
  \Big ( \big (\widehat{P_{P_n^{\vartheta^*}}(A)}\big )_1 \Big )}
  {\Big ( \big (\widehat{P_{P_n^{\vartheta^*}}(A)} \big )_N \Big )^2}.
\end{equation}
For simple sampling, one proceeds in the obvious analogous way
(without tilting and reweighting).

\begin{figure}[htbp]
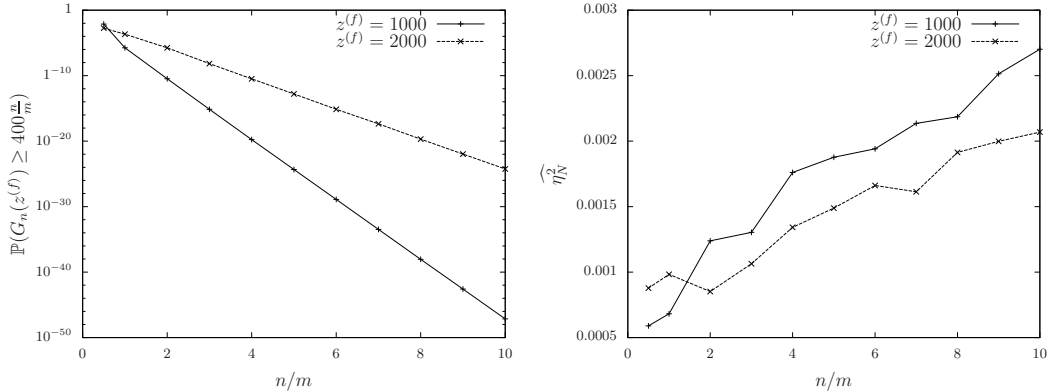


 \begin{minipage}[b]{7 cm}
    \scalebox{0.6}{\input{lim_probs_gact400.tex}}
  \end{minipage}
  \begin{minipage}[b]{7 cm}
    \scalebox{0.6}{\input{lim_res_gact400.tex}}
  \end{minipage}
  \caption{Importance sampling simulations  for
$\mathbb{P}(G_n(z^{(f)})\geq g_{\rm act} n/m)$ for $n=n_i$,
$0 \leq i \leq 10$, for $g_{\rm act}=400$ and two values of $z^{(f)}$.
Left: estimate of the probability (note that the vertical axis is on 
logarithmic scale). Right: estimated squared RE.}
  \label{fig:lim}
\end{figure}

In line with the limit discussed in Sec.~\ref{subsec:alg}, we
now consider $G_n(z^{(f)})$ for system sizes $n = n_i$, 
where $n_i=n_i^{(c)} + n_i^{(v)} + n_i^{(f)}$, $0 \leq i \leq 10$,
and we choose $n_i^{(\alpha)}= i m^{(\alpha)}$, $\alpha \in \{c,v,f\}$,
for  $1 \leq i \leq 10$, as well as   
$n_0^{(c)}=  m^{(c)}/2$,  $n_0^{(v)}=  m^{(v)}/2$, and  $n_0^{(f)}= m^{(f)}$
(i.e., we simply 
`multiply' the system, except for $i=0$, which corresponds to `half'
a system except for
the foreign peptide, which cannot be split into two). We then simulate
$\PP(G_{n_i}(z^{(f)}) \geq g_{\rm act} n_i/m)$ for two values of 
$z^{(f)}$ and a fixed value of $g_{\rm act}$
with our importance sampling method, as shown in Fig.~\ref{fig:lim}.

Obviously, the (estimated) probabilities decay to zero at an
exponential rate with increasing
$n$, as they must by their LDP. In contrast, the (estimated)
squared RE only increases linearly -- this even goes beyond the prediction
of the theory (asymptotic efficiency only guarantees a subexponential
increase).

\begin{figure}[htbp]
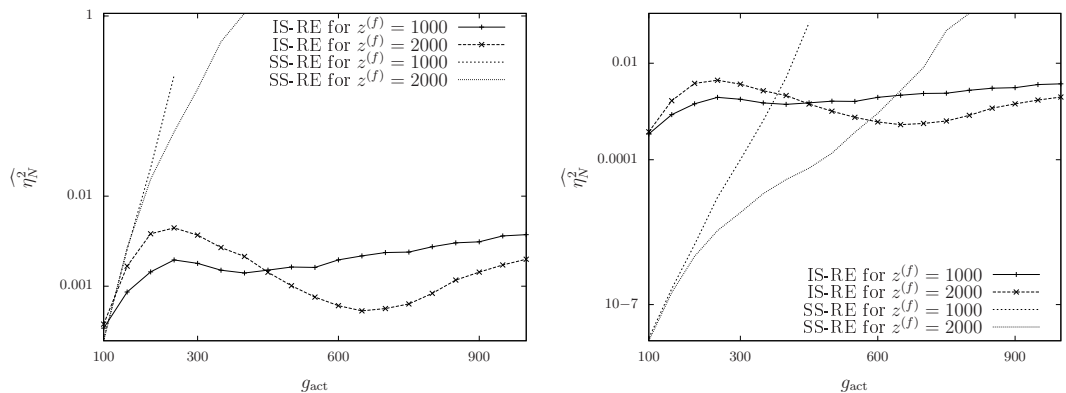


 \begin{minipage}[b]{7 cm}
    \scalebox{0.6}{\input{re_ss_is_less.tex}}
  \end{minipage}
  \begin{minipage}[b]{7 cm}
    \scalebox{0.6}{\input{re_ss_is.tex}}
  \end{minipage}
  \caption{Estimated squared RE for  simple sampling 
($N=10000$ times the number of steps contained in the considered interval
(left), 
$N=1.3*10^8$ (right)), and importance sampling 
($N=10000$ per threshold value in either panel) simulations of
$\mathbb{P}(G(z^{(f)})>g_{\rm act})$ of the basic model,
Eq.~\eqref{eq:the_model}.
Note that the vertical axis is on 
logarithmic scale.}
  \label{fig:re_is_ss}
\end{figure}

So far, we have considered the $n$-dependence of the method for
a fixed value of $g_{\rm act}$, in the light of the available
asymptotic theory. 
For the practical simulation of the given T-cell problem,
we now take the given system size $n=m$ and numerically investigate 
the relative error as a function of $g_{\rm act}$.
Here, the exponential decay of $\PP(G(z^{(f)}) \geq g_{\rm act})$
as a function of $g_{\rm act}$ is decisive, which we have already
observed in Fig.~\ref{fig:vergleich}, and which goes together with the
at-least-linear \emph{increase} of $I$ with $g_{\rm act}$ 
(recall that $I$ is convex, and see Fig.~\ref{fig:ratefunc}).
Fig.~\ref{fig:re_is_ss} shows the relative error of both SS and IS.
It does not come as a surprise that, again, IS does extremely
well and beats the exponential decay of the probabilities: whereas,
on the log scale of the vertical axis, the squared RE of SS grows
roughly linearly, it remains more or less constant for IS.
(The very low squared RE of the simple sampling graphs for low 
thresholds in the right panel is due to the fact that the probability to 
reach this threshold is quite high and the  huge sample
of $N=1.3 \cdot 10^8$ contributes to estimating it,
that is, the sample sizes are not comparable. A 
simple sampling simulation run with the total sample size of a corresponding
IS simulation (i.e., $N= 10000$ times the number of steps contained in the
interval considered) results 
in higher relative errors than for 
importance sampling even for the low 
threshold values (left panel). We would
like to note, however, that  the runtime of simple sampling for 
these small 
sample sizes is shorter than the runtime for IS, even if one does not count
the overhead required to get the tilting parameters for importance 
sampling.)

\begin{figure}[ht]
\input{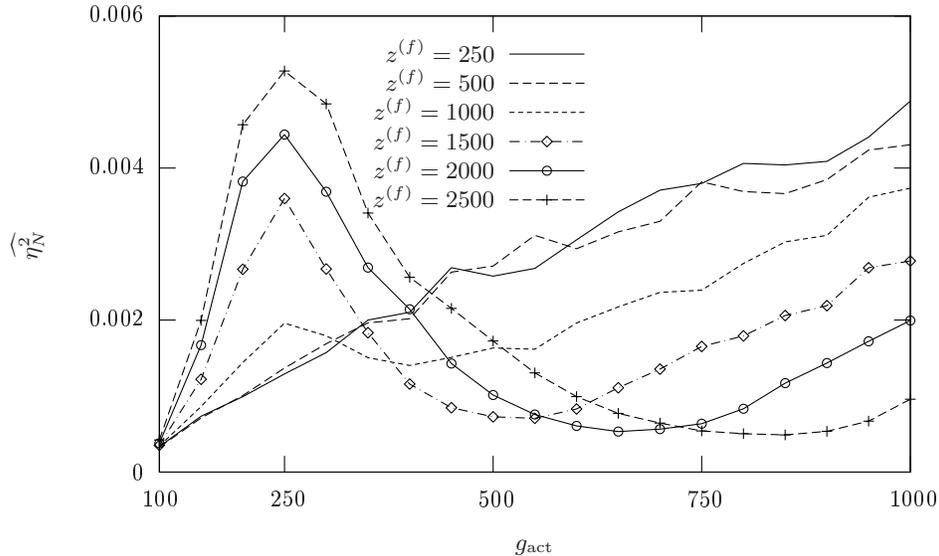}
\caption{Estimated squared RE of our IS estimate, for various
frequencies $z^{(f)}$ of the foreign antigen. Details are as in
Fig.~\ref{fig:re_is_ss}, but now the vertical axis is on linear scale.}
\label{fig:re_all_is}
\end{figure}
Figure \ref{fig:re_all_is} sheds  more light on the behaviour of the 
relative error of the IS simulation. It shows 
the squared RE for $6$ distinct $z^{(f)}$-values
 and  reveals the finite-size effects. The wave-like behaviour
for larger $z^{(f)}$ is due to the fact that, for very low 
threshold values, there is no real need for tilting, because
the original distribution $P_n$
is already  close to optimal and the tilting factor is very small. 
For increasing 
thresholds, substantial tilting is required, but 
there are still visible deviations from the $n \to \infty$ limit
(as already discussed  in the context of Fig.~\ref{fig:vergleich}),
so the tilted distributions are not optimal. This produces the
hump in the squared RE curves, which is more pronounced for
larger  $z^{(f)}$ values because, for the case $n=m$ considered here,
the foreign antigens come as a single term that may stand out.
For large $g_{\rm act}$, finally,
one gets close enough to the limit, and the expected sub-exponential
increase sets in (in our case, it is, in fact, roughly linear). 
Nevertheless, it should be clear that,
in spite of the slight non-optimality at small threshold values,
our tilted distributions still yield a far lower squared RE than does
simple sampling. A very similar picture emerges for the extended
model; surprisingly, the relative error is no larger than in the
basic model, although the \emph{ad hoc} simulation method used here
is not asymptotically efficient (see Sec.~\ref{subsec:ext_copy};
data not shown). 


\subsection{Analysis of the T-cell model}
In this Section, we  use our simulation method to obtain more detailed 
insight into the phenomenon of statistical recognition in the T-cell model. 
As discussed before, the task is to discriminate one foreign antigen type 
against a noisy background of a large number of self antigens. We already 
know from Fig. \ref{fig:vergleich} that, for threshold values that 
are not too small, the activation 
probability in the presence of foreign antigens is several orders of 
magnitude larger than the activation probability of the
self-background, i.e. Eq.~\eqref{eq:bed} 
is satisfied. As discussed in \cite{Zint:2008}, this 
distinction relies on $z^{(f)}>z^{(c)},z^{(v)}$ -- what
happens is that larger copy 
numbers of the foreign 
antigen thicken the tail of the distribution of $G(z^{(f)})$
(without changing its mean), so that 
the threshold is more easily surpassed. The self-nonself distinction 
may, according to this model, be roughly  described as follows. For a given 
antigen (foreign or self), finding a highly-stimulating T-cell receptor is 
a rare event; 
but if it occurs to a foreign antigen, it occurs many times simultaneously
since 
there are numerous copies, which all contribute the same large
signal, since  all receptors of the T-cell involved are 
identical; the resulting stimulation rate is thus high. In contrast,
if it is a self antigen that finds a highly-stimulating receptor, the effect
is less pronounced due to the smaller copy numbers. In this sense,
the toy model explains the distinction solely on the basis of
copy numbers; 
but see the Discussion for more sophisticated effects that alleviate
this requirement.

Following these intuitive arguments, we now aim at a more detailed 
picture of how the self background looks, and how 
the foreign type stands out against it.
To investigate this, it is useful to consider the histograms of the 
total constitutive, variable, and foreign stimulation rates, i.e., 
the contributions 
of the constitutive sum, the variable sum, and the
individual foreign term in the sum \eqref{eq:the_model}, 
either for all samples or for the subset of samples for which 
$ G(z_f)\geq g_{\rm act}$, 
for various $g_{\rm act}$.
Since this requires a 
higher resolution (and thus 
larger sample size) than the calculation of the activation probabilities 
alone, such analysis would be practically impossible with simple sampling.
With IS, we again generated
10000 samples per $g_{\rm act}$ value, from which between $30$ and 
$70$ percent turned out to reach the threshold. 

\begin{figure}[ht]
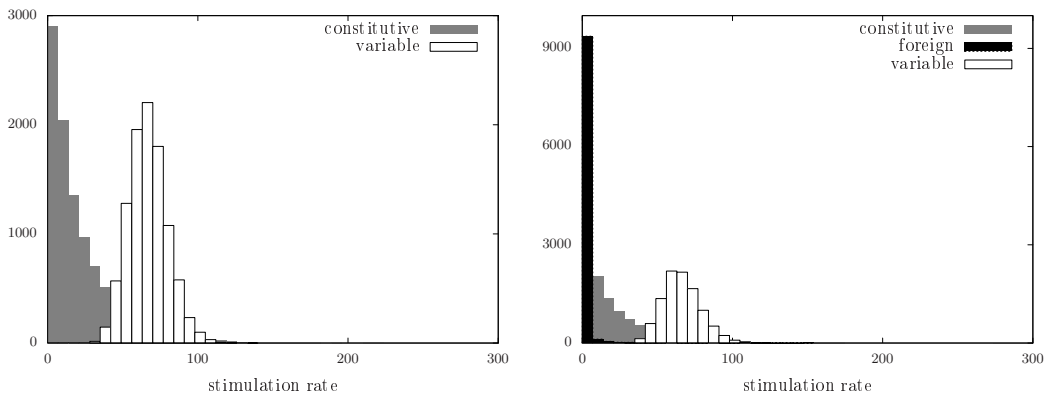


\begin{minipage}[b]{7 cm}
\scalebox{0.6}{\input{apost00_zf0_t0.tex}}
\end{minipage}
\begin{minipage}[b]{7 cm}
\scalebox{0.6}{\input{apost00_t0.tex}}
\end{minipage}

\caption{Histograms of the total stimulation rates  of variable, constitutive,
and foreign antigens, for $z^{(f)}=0$ (left) and $z^{(f)}=1000$ (right),
in the basic model \eqref{eq:the_model}, when all samples
are included. Sample size is 10000, 
and the vertical axis holds the number of samples whose total
constitutive (variable, foreign) stimulation rates fall into  
given intervals. Note that the
scaling of the vertical axis varies across diagrams.}
\label{fig:histo_t0}
\end{figure}

\begin{figure}[ht]
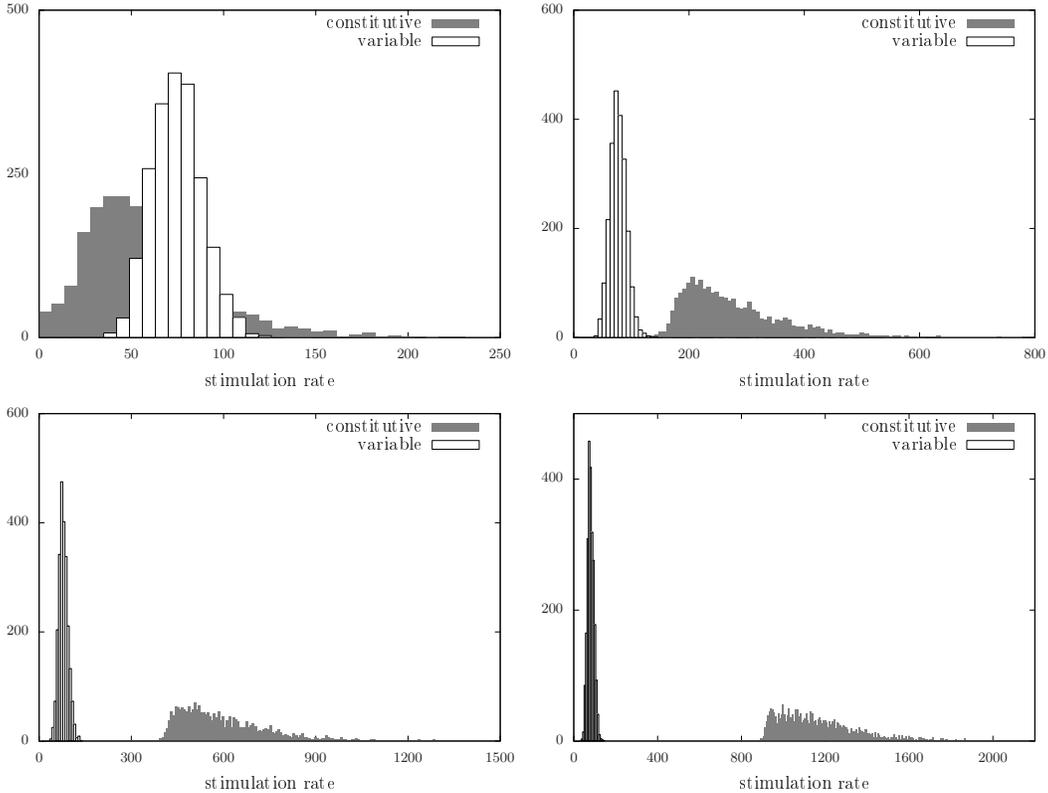


\begin{minipage}[b]{7 cm}
\scalebox{0.6}{\input{apost00_t1.tex}}
\end{minipage}
\begin{minipage}[b]{7 cm}
\scalebox{0.6}{\input{apost00_t4.tex}}
\end{minipage}

\begin{minipage}[b]{7 cm}
\scalebox{0.6}{\input{apost00_t9.tex}}
\end{minipage}
\begin{minipage}[b]{7 cm}
\scalebox{0.6}{\input{apost00_t19.tex}}
\end{minipage}
\caption{Histograms of the total stimulation rates  of variable and constitutive
antigens, for $z^{(f)}=0$, in the basic model \eqref{eq:the_model},
for samples that reach a given threshold value 
($g_{\rm act}=100$ (upper left), $g_{\rm act}=250$ (upper right), 
$g_{\rm act}=500$ (lower left), $g_{\rm act}=1000$ (lower right)).
Sample size is 10000, and the vertical axis holds the number of
samples  that reach   $g_{\rm act}$ and  whose total
constitutive (variable, foreign) stimulation rates falls into  
given intervals.
Note that the
scaling of both axes varies across diagrams.}
\label{fig:histo2}
\end{figure}

\begin{figure}[ht]
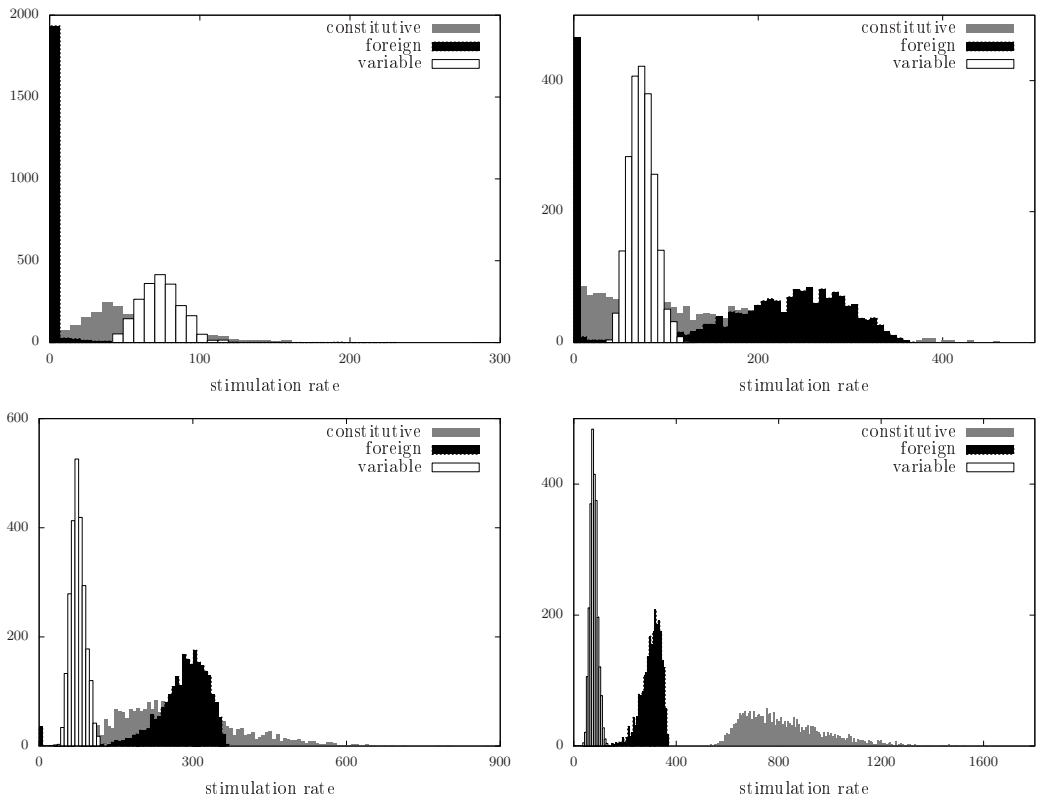


\begin{minipage}[b]{7 cm}
\scalebox{0.6}{\input{apost1000_t1.tex}}
\end{minipage}
\begin{minipage}[b]{7 cm}
\scalebox{0.6}{\input{apost1000_t4.tex}}
\end{minipage}

\begin{minipage}[b]{7 cm}
\scalebox{0.6}{\input{apost1000_t9.tex}}
\end{minipage}
\begin{minipage}[b]{7 cm}
\scalebox{0.6}{\input{apost1000_t19.tex}}
\end{minipage}
\caption{Histograms of the total constitutive, variable and foreign
stimulation rates for  $z^{(f)}=1000$
in the basic model \eqref{eq:the_model}.
Sample size is 10000, and the vertical axis holds the number of
samples  that reach the threshold  $g_{\rm act}$ and  whose total
constitutive (variable, foreign) stimulation rate falls into a 
given interval, for
$g_{\rm act}=100$ (upper left), $g_{\rm act}=250$ (upper right), 
$g_{\rm act}=500$ (lower left), $g_{\rm act}=1000$ (lower right). 
The  maximal stimulation rate for the foreign antigens  is
$z^{(f)} w(1)=367.9$. Note that the
scaling of both axes varies across diagrams.}
\label{fig:histo}
\end{figure}

\begin{table}
\begin{minipage}{7cm}
$$
\begin{array}{|c|c|c|c|c|c|}
\hline
{\rm rate}  \setminus g_{\rm act} & 0 & 100 & 250 & 500 & 1000 \\
\hline
{\rm variable} & 66.6 & 74.9 & 77.1 & 78.8 & 80.0 \\
\hline
{\rm constitutive} & 22.2& 59.2 & 277.7 & 590.6& 1160 \\
\hline
\end{array}
$$
\end{minipage}
\begin{minipage}{7cm}
$$
\begin{array}{|c|c|c|c|c|c|}
\hline
{\rm rate}  \setminus g_{\rm act}  & 0 & 100 & 250 & 500 & 1000 \\
\hline
{\rm variable} & 12.7 & 13.9 &14.5 & 14.9 & 15.1 \\
\hline
{\rm constitutive} & 23.1 & 35.6 & 88.8&134.9 & 191.3 \\
\hline
\end{array}
$$
\end{minipage}
\caption{Sample means (left) and sample standard deviations (right)
of the histograms in Fig.~\ref{fig:histo_t0} (left) and Fig.~\ref{fig:histo2}
(i.e., the self-only case).}
\label{tab:without_foreign}
\end{table}

\begin{table}
\begin{minipage}{7cm}
$$
\begin{array}{|c|c|c|c|c|c|}
\hline
{\rm rate}  \setminus g_{\rm act} & 0 & 100 & 250 & 500 & 1000 \\
\hline
{\rm variable} & 65.9 &  74.1 & 74.2 & 76.2 & 78.4 \\
\hline
{\rm constitutive} & 21.8 & 55.9 & 129.5 & 270.4& 821.1\\
\hline
{\rm foreign} & 0.9 & 4.0 & 184.8& 279.6 & 302.2 \\
\hline
\end{array}
$$
\end{minipage}
\begin{minipage}{7cm}
$$
\begin{array}{|c|c|c|c|c|c|}
\hline
{\rm rate}  \setminus g_{\rm act}  & 0 &  100 & 250 & 500 & 1000 \\
\hline
{\rm variable} & 12.7 & 14.1 & 13.9 & 14.2 & 14.7 \\
\hline
{\rm constitutive} & 22.4 & 42.0 & 90.4& 109.1& 163.7 \\
\hline
{\rm foreign} & 6.7& 18.5 & 112.2 &  54.5& 39.2 \\
\hline
\end{array}
$$
\end{minipage}
\caption{Sample means (left) and sample standard deviations (right)
of the histograms in Fig.~\ref{fig:histo_t0} (right) and Fig.~\ref{fig:histo}
(i.e., the case with foreign antigens).}
\label{tab:foreign}
\end{table}

\begin{figure}[ht]
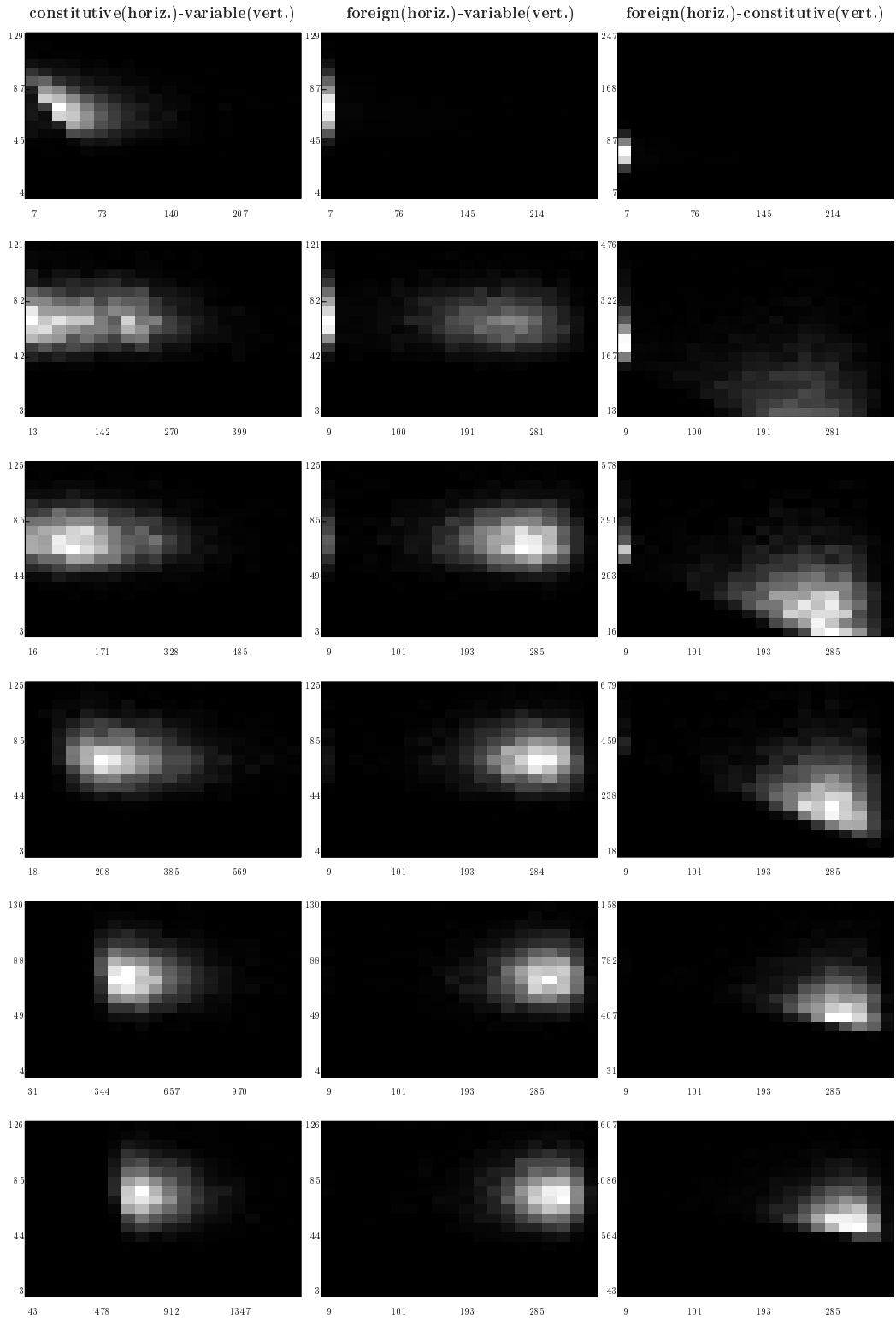


\begin{minipage}[b]{4.5 cm}
\scalebox{0.5}{\input{graymap_t1_a1a2.tex}}
\end{minipage}
\begin{minipage}[b]{4.5 cm}
\scalebox{0.5}{\input{graymap_t1_a1a3.tex}}
\end{minipage}
\begin{minipage}[b]{4.5 cm}
\scalebox{0.5}{\input{graymap_t1_a2a3.tex}}
\end{minipage}

\begin{minipage}[b]{4.5 cm}
\scalebox{0.5}{\input{graymap_t4_a1a2.tex}}
\end{minipage}
\begin{minipage}[b]{4.5 cm}
\scalebox{0.5}{\input{graymap_t4_a1a3.tex}}
\end{minipage}
\begin{minipage}[b]{4.5 cm}
\scalebox{0.5}{\input{graymap_t4_a2a3.tex}}
\end{minipage}

\begin{minipage}[b]{4.5 cm}
\scalebox{0.5}{\input{graymap_t6_a1a2.tex}}
\end{minipage}
\begin{minipage}[b]{4.5 cm}
\scalebox{0.5}{\input{graymap_t6_a1a3.tex}}
\end{minipage}
\begin{minipage}[b]{4.5 cm}
\scalebox{0.5}{\input{graymap_t6_a2a3.tex}}
\end{minipage}

\begin{minipage}[b]{4.5 cm}
\scalebox{0.5}{\input{graymap_t9_a1a2.tex}}
\end{minipage}
\begin{minipage}[b]{4.5 cm}
\scalebox{0.5}{\input{graymap_t9_a1a3.tex}}
\end{minipage}
\begin{minipage}[b]{4.5 cm}
\scalebox{0.5}{\input{graymap_t9_a2a3.tex}}
\end{minipage}

\begin{minipage}[b]{4.5 cm}
\scalebox{0.5}{\input{graymap_t14_a1a2.tex}}
\end{minipage}
\begin{minipage}[b]{4.5 cm}
\scalebox{0.5}{\input{graymap_t14_a1a3.tex}}
\end{minipage}
\begin{minipage}[b]{4.5 cm}
\scalebox{0.5}{\input{graymap_t14_a2a3.tex}}
\end{minipage}

\begin{minipage}[b]{4.5 cm}
\scalebox{0.5}{\input{graymap_t19_a1a2.tex}}
\end{minipage}
\begin{minipage}[b]{4.5 cm}
\scalebox{0.5}{\input{graymap_t19_a1a3.tex}}
\end{minipage}
\begin{minipage}[b]{4.5 cm}
\scalebox{0.5}{\input{graymap_t19_a2a3.tex}}
\end{minipage}

\caption{Pairwise joint frequencies of the total constitutive, variable,
and foreign stimulation rates, for those samples with
$G(z^{(f)})>g_{\rm act}$ in the basic model \eqref{eq:the_model}
(with $z^{(f)}=1000$).  Greyscales correspond to
number  of samples 
falling into  2D-intervals defined by total stimulation rates of pairs of  
antigen types. Rows (from top to bottom): 
$g_{\rm act}=100, 250, 350, 500, 750,1000$; columns (from left to right):
constitutive (horizontal) -- variable (vertical); 
foreign (horizontal) -- variable (vertical); 
foreign (horizontal) -- constitutive (vertical). Lighter
shading corresponds to higher frequencies.}
\label{fig:graymaps}
\end{figure}

Figure \ref{fig:histo_t0} shows the resulting histograms
when all samples are included, 
and  Figs.~\ref{fig:histo2}  
and \ref{fig:histo} show the  histograms for
the subset of samples that have surpassed four representative threshold
values, without and with foreign 
antigen. Tables 
\ref{tab:without_foreign} and \ref{tab:foreign} summarise these 
results in terms of means and standard deviations.
Finally, Fig.~\ref{fig:graymaps} 
shows the 
corresponding two-dimensional statistics
for all pairs of variable, constitutive, and 
foreign stimulation rates, again for various threshold values. 
(Figs.~\ref{fig:histo2}--\ref{fig:graymaps} are based on 
the outcome of importance sampling
\emph{without reweighting};
normalising by the number of "successful"
samples would result in an estimate of the conditional distribution,
because the reweighting factors cancel out.)

Let us start with the situation without foreign antigens, as displayed in 
Figs.~\ref{fig:histo_t0} (left) and \ref{fig:histo2} as well as
Table~\ref{tab:without_foreign}. This already illustrates the fundamental 
difference between variable and constitutive antigens. Judging from the 
large number ($m^{(v)} = 1500$) of individual terms in the sum
at low copy number ($z^{(v)}=50$),
the variable stimulation
rate is expected to be approximately normally distributed and fairly closely 
peaked around its mean -- 
at least as long as no restriction on $G(z^{(f)})$ is involved -- and, 
as the Figure 
shows, this feature persists when $G(z^{(f)})>g_{\rm act}$,
practically independently of 
the threshold involved. So, the variable antigens form a kind of background 
that poses no difficulty to foreign-self distinction: it is not very noisy, 
and it does not change with the threshold.

In contrast, the distribution of the constitutive activation rates
is wider; this is due to the large copy numbers
($z^{(c)}=500$), the effect of which is not compensated by the
smaller number of terms, $m^{(c)}=50$. Furthermore, the
normal approximation is not expected to be particularly 
good for the constitutive antigens -- given the extreme asymmetry of the 
$W$-distribution (see Fig.~\ref{fig:w}), the central 
limit theorem will not average out 
the deviations at only $m^{(c)}=50$. 
In particular, the distribution remains asymmetric. 
With increasing threshold, this distribution 
moves to the right. 
The reason for this is that, in order  to reach an increasing
$g_{\rm act}$, the  
tail events of the constitutive
or the variable sum or both must be used, but it is ``easier'' (that is, more probable)
to use the constitutive one because it contains more atypical events.
In the language of large deviation 
theory, this is an example of the general principle that ``large deviations are 
always done in the the least unlikely of all the unlikely ways''
\cite[Ch.~I]{Hollander:2000}. In the language of
biology, the constitutive antigens are the problem of foreign-self 
distinction: due to their high copy numbers and incomplete averaging, 
fluctuations persist that occasionally induce an immune response even in 
the absence of foreign antigens. This occurs if a T-cell receptor happens 
to fit particularly well to one, or a number of,  constitutive antigen 
types on an APC; 
due to their large copy numbers, these few highly-stimulating types are
then sufficient to surpass the threshold (in contrast, several
highly-stimulating types would be required for the variable antigens to elicit 
a reaction, which is too improbable).

Let us now turn to the picture with foreign antigen present 
(Figs.~\ref{fig:histo_t0} (right), \ref{fig:histo},  \ref{fig:graymaps},
and Table~\ref{tab:foreign}). One salient 
feature here is that the variable stimulation rate behaves exactly as in the 
self-only case: closely peaked around a small mean, unchanged 
when $\{G(z^{(f)})>g_{\rm act}\}$ is imposed. 
The picture is thus dominated by the interplay of constitutive and foreign types. 
In line with Fig. \ref{fig:vergleich}, the situation is similar in the
case without restriction on  $G(z^{(f)})$ (Fig.~\ref{fig:histo_t0}, right)
and the case when $G(z^{(f)})\geq100$ 
(Fig.~\ref{fig:histo}, upper left). In particular, the foreign 
stimulation rate is 
closely peaked at $0$; only the constitutive background has moved slightly to the right, 
exactly as in the self-only case. For $g_{\rm act}=250$ 
(Fig.~\ref{fig:histo}, upper right), where, according to 
Fig. \ref{fig:vergleich}, foreign-self distinction sets in, the foreign 
stimulation rate  becomes prominent: the right branch of the $W$-distribution 
now becomes populated, and the associated stimulation rates are large due 
to the large copy numbers $z^{(f)}$ involved. 

Nevertheless, for $g_{\rm act}=250$, the foreign stimulation rate is  close to $0$ in a 
sizable fraction of the cases in which an immune reaction occurs -- here, 
the reaction is brought about by the constitutive background, which moves to 
the right just as in the self-only case (but less pronounced).
Fig. \ref{fig:graymaps} shows that the constitutive and foreign stimulation rates 
are, indeed, negatively correlated: as is to be expected, low foreign rates 
are compensated by high constitutive rates and vice versa
(in contrast, the 
variable background hardly correlates with either the
constitutive or the foreign stimulation rate). As in the 
self-only 
case, therefore, the level of unwanted activation 
(``self-only'' or ``mainly self, without appreciable foreign activation'')
is set by the tail behaviour of the constitutive background. However, if 
$g_{\rm act}$ is increased further (Fig.~\ref{fig:histo}, lower left), every T 
cell beyond the threshold displays high stimuli for the foreign antigen, 
their distribution shifting  even further to the right and concentrating near the 
maximal stimulation rate given by the maximum of the function $w$ of 
Eq.~\eqref{eq:transformation_function}, more precisely,
by $z^{(f)} w(1)$. This maximum can, of course, not change 
by imposing restrictions on $G(z^{(f)})$; thus, any further increase of $g_{\rm act}$
(Fig.~\ref{fig:histo}, lower right) must then
be matched by the by now familiar shift of the constitutive background.
(This last panel is, however,  less biologically realistic
since the probabilities involved are too small to be relevant -- after all,
with about $10^7$ different T-cell types, threshold values that yield
activation probabilities far below $10^{-7}$ even in the presence of foreign
antigens offer no immune protection.)

\begin{figure}[ht]
\scalebox{0.75}{\input{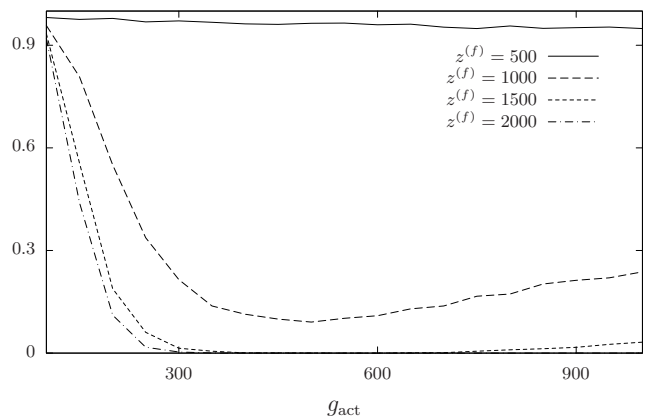}}
\caption{Fraction of samples whose self-component alone
is above threshold,
among those that reach the threshold
in the presence of $z^{(f)}$ foreign molecules, for various $z^{(f)}$
(i.e., IS simulation of the probability
in Eq.~\eqref{eq:cond_react}).  
Sample size is 10000 for each $g_{\rm act}$ value considered.}
\label{fig:reaching_ratios}
\end{figure}

A further illustration of the onset of self-nonself
distinction is presented in Fig. \ref{fig:reaching_ratios}. Here we
consider 
\begin{equation}\label{eq:cond_react}
\mathbb{P}\big (G(z^{(f)})- z^{(f)}W_{n^{(c)}+n^{(v)}+1}>g_{\rm act} \mid
G(z^{(f)}) > g_{\rm act} \big ) 
= \frac{\mathbb{P} \big ( G(z^{(f)})- z^{(f)}W_{n^{(c)}+n^{(v)}+1}>g_{\rm act} \big )}
  {\mathbb{P} \big (G(z^{(f)})>g_{\rm act} \big )},
\end{equation}
i.e., the probability that, in a T-cell that is activated in the presence
of foreign antigen, the self component alone would have been sufficient
for the activation.
From $z^{(f)}=1000$ onwards, this probability decreases to 0 quickly
with increasing $g_{\rm act}$. Put
differently, in large parameter regions, the foreign antigens do indeed
make the difference, which is the decisive feature of self-nonself
distinction.

\section{Conclusion and outlook}
We have established here a method of LD sampling that allows
the convenient simulation of the rare events relevant to statistical 
recognition in the immune system. Thus a more thorough investigation
of these events could  be carried out.

But this is only a first step, and the
goal for future work is to use this or related methods to investigate
biologically realistic models. Indeed, the toy model
considered here, which relies solely on distinction by copy numbers,
does serve the aim to illustrate that distinction
against a noisy background is, at all, possible, even without an
intrinsic difference between self and nonself, and how this is
related to the rare events in the tail of the background distribution.
However, biologically realistic models have to take into account
tolerisation mechanisms that make the T-cells less responsive to
self antigens. One important such mechanism is
so-called \emph{negative selection}.
Negative selection occurs during the maturation phase of young
T-cells in the thymus, before they are released into the body.
In a process similar to the one described by the toy model,
they are confronted with APCs that present mixtures of
various self antigens, and those T-cells whose activation rate
surpasses a thymic activation threshold $g_{\rm thy} < g_{\rm act}$ are
eliminated. When they are later, after leaving the thymus, confronted
with mixtures of self and foreign antigens, the stimulation rates
emerging from self and foreign are no longer i.i.d. (the self ones
are biased towards smaller values and possibly  negatively correlated). 
In fact, a simple model for negative selection was already
described in BRB \cite{Berg:2001}, and shown to  
drastically reduce the self background,
so that  foreign antigens do no longer require elevated copy numbers
to be detected.
More sophisticated models of negative selection have been formulated e.g.\
in \cite{Berg:2003}. However, their simulation still awaits the
development of adequate methods.
This is the purpose of ongoing work.

\section{Acknowledgements}
It is our pleasure to thank Michael Baake and Natali Zint for critically
reading the manuscript, and Hugo van den Berg and Frank den Hollander 
for helpful discussions. This work was supported by DFG-FOR 498
(Dutch-German Bilateral Research Group on 
Mathematics of Random Spatial Models in Physics and Biology) and
the NRW International Graduate School of Bioinformatics and Genome
Research at Bielefeld University.

\clearpage


\end{document}